\documentclass[%
preprint,
onecolumn,
titlepage,
tightenlines,
12pt
]
{revtex4-2}

\usepackage{dcolumn}

\usepackage{tikz}
\usepackage{amsmath, bm, amssymb,graphicx}

\usepackage{subcaption}
\usepackage{caption}

\usepackage[export]{adjustbox}

\usepackage{xcolor}
\definecolor{myBlue}{rgb}{0.12, 0.51, 0.75}

\newcommand{\greenline}{\raisebox{2pt}{\tikz{\draw[-,black!30!green,dashed,line width = 1pt](0,0) -- (5mm,0);}}}

\newcommand{\baseline}{\raisebox{2pt}{\tikz{\draw[-,orange,dashed,line width = 1pt](0,0) -- (5mm,0);}}}

\newcommand{\greyline}{\raisebox{2pt}{\tikz{\draw[-,gray,solid,line width = 0.9pt](0,0) -- (5mm,0);}}}

\newcommand{\blueline}{\raisebox{2pt}{\tikz{\draw[-,myBlue,solid,line width = 0.9pt](0,0) -- (5mm,0);}}}

\graphicspath{{Figures/}}

\begin{document}

\preprint{APS/123-QED}

\title{Surface morphing for aerodynamic flows at low and stalled angles of attack}% Force line breaks with \\

\author{Ernold Thompson}
 \email{ernoldt2@illinois.edu}
\affiliation{%
 Department of Aerospace Engineering, University of Illinois at Urbana-Champaign, IL 61801, USA
}%

\author{Andres Goza}%
\affiliation{%
 Department of Aerospace Engineering, University of Illinois at Urbana-Champaign, IL 61801, USA
}%


\begin{abstract}
In the current work we numerically study the effect of traveling-wave surface morphing actuation, which is a lightweight, spatially distributed actuation strategy made possible by advances in materials science. 
Although this actuation strategy has been studied at higher Reynolds numbers for an airfoil and a rectangular flat plate, its effects at low Reynolds numbers relevant to micro-air vehicles have not yet been investigated.
We perform high-fidelity 2D numerical simulations to study the effects of traveling wave surface morphing on the suction surface of NACA0012 at $Re=1{,}000$. The kinematics of actuation are defined by wavenumber and wavespeed, both of which are varied over a wide range of values to include parameters that considerably change the lift dynamics as well as those that do not. We first study the effect of actuation at an angle of attack of $\alpha = 5^{\circ}$, where the unactuated flow is steady. The lift dynamics are found to align with the surface morphing kinematics, and there is a low-pressure minimum shown to be introduced into the flow-field by morphing that advects at a speed agnostic to the morphing parameters. Lift benefits are found to be maximal when the morphing kinematics align with this intrinsic flow speed. We then investigate the role of morphing in the presence of an unsteady, separated baseline flow (with intrinsic vortex-shedding processes) at $\alpha=15^\circ$. At this higher angle of attack, we identify three distinct behavioral regimes based on the relationship between morphing and the underlying shedding frequency. Of these regimes, the most beneficial to mean lift is the lock-on regime, where the vortex-shedding dynamics align with the morphing kinematics. Lock-on was similarly found using this actuation strategy at higher Reynolds numbers, though in that setting the effect was to reduce separation, whereas at these lower Reynolds numbers the outcome is that vortex shedding persists with---in certain cases---significant lift benefits. We also identify other regimes where morphing can become out of phase with the vortex-shedding dynamics, termed here the interactive regime, and where morphing leaves the unactuated dynamics unaltered, termed here the superposition regime. At the higher angle of attack, parameters leading to lift benefits/detriments are explained in terms of the effect of morphing on the leading and trailing-edge vortex. Where appropriate, connections between the mechanisms at the higher angle of attack are drawn to the matching/disparity of timescales between morphing and lift-producing pressure signatures seen in the lower-angle-of-attack setting. 
\end{abstract}
\maketitle
\section{Introduction}
Control of low-Reynolds-number flows past aerodynamic bodies could lead to the design of next generation micro- and unmanned-aerial vehicles. This control aim is challenging because of the associated complex, unsteady, and nonlinear flow dynamics. Moreover, for this control imperative to be successful, fast actuation strategies capable of responding to the unsteady and nonlinear flow dynamics are necessary. Because of the intrinsic complexity of these aerodynamic flows, actuation strategies are often first developed within the setting of a stationary airfoil, and subsequently extended to account for more complex bodies and unsteady kinematics. We restrict ourselves to this stationary-airfoil setting in this article.

Active flow control strategies involving synthetic jet \citep{amitay2001aerodynamic,Glezer2002, raju2008dynamics, Agashe2009}, combustion \citep{Glezer2003,Cutler2005,Crittenden2009}, and plasma \citep{Corke2007,Corke2010,Moreau2007} actuation have been used to modulate the near-body flow and provide lift and/or drag benefits. An alternative to these approaches, leveraging advances in materials science, is surface morphing actuation. This actuation strategy imposes small-amplitude oscillations along the airfoil surface, and utilizes lightweight devices while simultaneously enabling spatially distributed actuation that can be driven across a wide range of timescales. \citet{munday2002active} implemented this actuation approach by embedding a flexible piezoelectric actuator into the airfoil suction surface and driving deformations of the airfoil surface via electrical actuation. The authors found that dynamic deformations yielded a reduction in flow separation. 

Since this study, various efforts have probed the effect of periodically driven surface morphing on aerodynamic performance. In some of these studies, the periodic excitations were prescribed as standing waves. \citet{jones2016exp} experimentally studied standing-wave morphing at a Reynolds number of $Re=50,000$ for a variety of temporal frequencies and angles of attack. The authors found increased velocity within the time-averaged boundary layer for certain morphing frequencies, and an associated increase in mean lift and reduction in mean drag. The aerodynamic benefits were maximal when the morphing frequency matched the dominant frequency associated with vortex shedding of the unactuated case. \citet{jones2018numerical} subsequently performed simulations at the same Reynolds number for an airfoil at an angle of attack of $\alpha=0^\circ$, and select morphing frequencies based on the fundamental shear layer frequency of the unactuated flow. This collection of morphing frequencies was found to improve mean lift, and a dynamic mode decomposition \citep{schmid2010dynamic,rowley2009spectral} demonstrated that at these specific frequencies shear layer roll-up occurred further upstream than in the unactuated case, yielding a corresponding increase in momentum transfer within the boundary layer and decrease in flow separation. The early shear-layer roll-up was shown to coincide with vorticity generation on the airfoil surface that was synchronized with the morphing frequency, leading the authors to argue that the aerodynamic benefits were associated with a \emph{lock-on} 
phenomenon where key flow dynamics were synchronized with the driving actuation. 

Other efforts have considered the use of travelling wave surface morphing. \citet{akbarzadeh2019numerical} numerically studied downstream-travelling morphing waves for an airfoil at $Re=50{,}000$ and an angle of attack of $\alpha=10^\circ$, for different actuation frequencies and amplitudes. At a fixed actuation amplitude, the mean lift was found to increase with actuation frequency until a threshold frequency was reached, at which point there were negligible mean lift improvements. The authors argued that the lift enhancement mechanism was an increase in momentum transfer within the boundary layer, similar to what was found for standing-wave actuation in \citet{jones2018numerical}. Increasing the amplitude eventually produced adverse effects, with increased flow separation and associated reductions in mean lift. In a subsequent study, \citet{akbarzadeh2019reducing} numerically investigated the effect of downstream travelling waves on flow over a flat plate at an angle of attack of $\alpha=10^\circ$ for three different actuation frequencies and six distinct spatial wavelengths. The use of these two parameters allowed for separate means by which to tune the morphing wave speed. Improvements in mean lift and drag with increasing frequency were demonstrated, but there was a non-monotonic dependence of these performance metrics on the wavelength. The authors used these results to argue that wavespeed unto itself did not dictate the ensuing flow dynamics, and that the frequency and wavelength of morphing had distinct effects on the flow. 

In this article, we use two-dimensional high-fidelity simulations involving a NACA0012 airfoil at a low Reynolds number of $Re=1{,}000$ to build on the above literature. First, there are (to our knowledge) no investigations of this spatially distributed surface morphing actuation framework for low Reynolds numbers relevant to micro-air vehicle flight. Second, the aforementioned studies have predominately been at lower angles of attack. As such, the effectiveness and driving physical mechanisms of surface morphing in the presence of unsteady flow separation---which becomes more pronounced at lower Reynolds numbers where the lower flow inertia is more susceptible to an adverse pressure gradient---are not known. For example, it is unknown whether lock-on mechanisms continue to produce aerodynamic benefits, and whether these mechanisms manifest themselves differently in strongly separated aerodynamic flows. Finally, the prior investigations have considered a small number of surface morphing frequencies and wavelengths, and a more general parametric dependence of aerodynamic performance on these parameters has not been clarified. For example, it is unclear whether frequencies with non-harmonic relationships to the unactuated vortex shedding processes produce aerodynamic benefits, and if so, by what mechanisms.

We focus here on surface morphing prescribed as a traveling wave for systematically varied wavespeeds
and wavenumbers. We consider two angles of attack: $\alpha=5^\circ$ and $\alpha=15^\circ$. At this low Reynolds number, the lower angle of attack yields a steady flow, which allows the effect of surface morphing in the absence of intrinsic flow frequencies to be clarified. 
At the larger angle of attack, the flow is unsteady and the associated canonical vortex shedding process leads to a von-Karman vortex street in the airfoil's wake. We quantify the changes in aerodynamic performance for the two angles of attack as a function of the morphing wavespeed and wavenumber, and characterize the interplay between the morphing and flow dynamics in the steady and unsteady separated flow regimes, drawing connections to the higher Reynolds number cases as appropriate. 

We emphasize that a broader class of surface-deformation based actuators have been developed for active flow control (see, e.g., \citet{wiltse1993manipulation,Seifert1998,Jeon2000,kang2015effects} and the review of \citet{cattafesta2011actuators}). As such, mechanistic insights into the surface morphing framework considered here could inform actuation protocols in these other settings. (Indeed, investigations of other surface-driven actuation strategies have reported aerodynamic performance benefits through lock-on effects \citep{kang2015effects}, suggesting similar mechanisms prevail across these actuation technologies). Similarly, passively compliant actuators have shown potential utility in aerodynamic flow control \citep{lei2014unsteady,curet2014aerodynamic,hussein2015flow,bohnker2019control,barnes2021initial}, and insights into the desired structural dynamics obtained via prescribed motion could provide a means to back out material properties that would induce the desired fluid-structure interplay.

\section{Problem setup and numerical methodology}
\subsection{Problem setup}

\begin{figure}
    \centering
    \includegraphics[width=0.55\textwidth, height=0.28\textwidth]{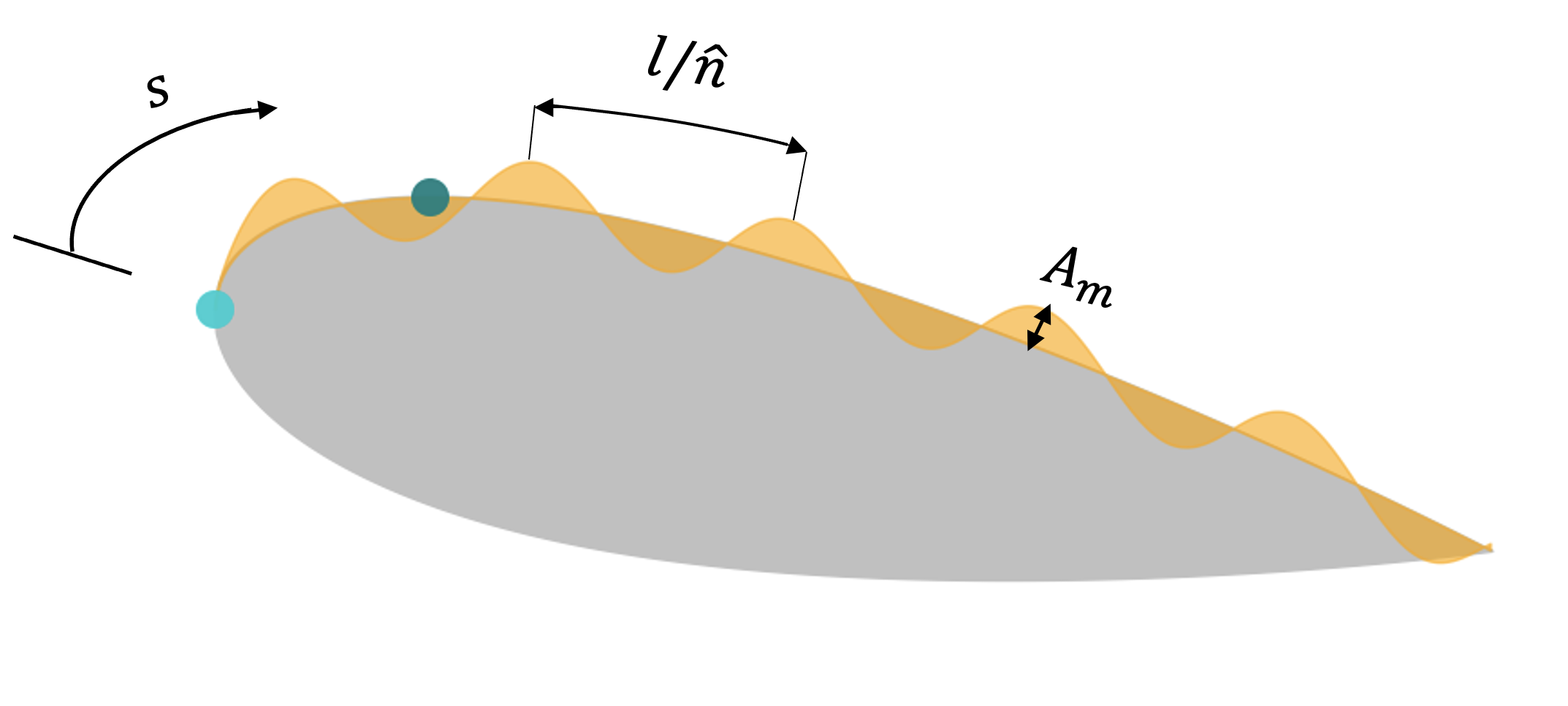}
    \caption{Schematic of prescribed surface morphing}
    \label{fig:Schematics}
\end{figure}

The prescribed surface morphing is applied to a NACA-0012 airfoil at a fixed  angle of attack, $\alpha$, with a Reynolds number of $Re = 1{,}000$. The surface morphing is applied normal to the airfoil surface in the form of a traveling wave, and is modelled as a velocity boundary condition applied at the undeformed suction surface of the airfoil. This representation of the surface actuation is rooted in the observation that surface deformations caused by morphing are small, so that the nonzero velocity drives the changes in the flow dynamics. The normal velocity profile at the suction surface, $v_n$, may be expressed as a function of the arc-length of a point on the suction surface from the leading edge, $s$, as
\begin{equation}
    v_n (s,t) = A_m \cos\left[2 \pi n (s-wt)\right],
    \label{eqn:normal_vel}
\end{equation}
where $A_m$ is the amplitude of the prescribed surface velocity, and $w$ and $n$ are respectively the wavespeed and wavenumber associated with the traveling wave. See figure \ref{fig:Schematics} for a schematic of the prescribed actuation. All quantities are presented in dimensionless form, normalized by the chord length $c$. Note that this nondimensionalization means the arclength at the trailing edge, $l$ say, is not unity because of the nonzero airfoil thickness. As such, the actual number of waves along the airfoil surface is $\hat{n}=ln/c$. We use the scaled wavenumber $n$ for simplicity to convey all results below, keeping in mind that it incorporates the nonunity factor $l/c$.

To be consistent with the small-amplitude nature of surface morphing actuation and focus on the effect of wavenumber and wavespeed, the morphing amplitude $A_m$ is fixed at $0.03$. Appreciable changes were not found for morphing amplitudes between $A_m \in[0.01, 0.05]$. Morphing amplitude of lower orders of magnitude did not result in observable changes from the unactuated case while amplitudes of higher magnitude would invalidate the small deformations inherent to surface morphing. The wavespeeds and wavenumbers considered in this work are $w \in [0.1,1]$ and $n \in [1,6]$, respectively. These ranges are selected to encompass the characteristic speed and lengthscale associated with the fundamental vortex shedding processes in the unactuated case at an angle of attack of $\alpha=15^\circ$. As such, the results below will probe the effect of surface morphing when the morphing parameters are smaller than, matched with, and larger than the intrinsic flow speeds and lengthscales. Note that the wavenumber and wavespeed induce a morphing frequency, $f_m=wn$ and an associated morphing period $T_m=1/f_m$.

Analysis of the aerodynamic system will often be done in terms of the coefficients of pressure and sectional lift and drag, defined respectively here as
\begin{equation}
    C_p=\frac{p-p_\infty}{\frac{1}{2}\rho U_\infty^2}, \quad C_l = \frac{F_y}{\frac{1}{2}\rho U_\infty^2 c}, \quad C_d = \frac{F_x}{\frac{1}{2}\rho U_\infty^2 c},
\end{equation}
where $\rho$ is the fluid density, $U_\infty$ is the freestream flow speed, $p$ is the pressure field variable, $p_\infty$ is a freestream pressure reference value, and $F_x$ and $F_y$ are the dimensional integrated force along the airfoil surface in the horizontal (streamwise) and vertical directions, respectively. 

\subsection{Numerical methodology}

The simulations are performed using the immersed boundary projection method as described in \citet{colonius2008fast}. The method solves the incompressible Navier-Stokes equations with applied forcing to satisfy the velocity boundary condition on the surface of the body, written here in dimensionless form as
\begin{eqnarray}
\frac{\partial \bm{u}}{\partial t} + \bm{u} \cdot \nabla \bm{u}  &=& -\nabla p + \frac{1}{Re} \nabla^2 \bm{u} + \int_{\Gamma} \bm{f}_b(\bm{\xi}) \delta (\bm{\xi} -\bm{x}) d\bm{\xi} \label{eqn:NS} \\
\nabla \cdot \bm{u} &=&0 \\
\int_{\Omega} \bm{u}(\bm{x}) \delta (\bm{x}-\bm{\xi}) d\bm{x} &=&  \bm{u}_b(\xi)\label{eqn:BC}
\end{eqnarray}
In the above equations, $\bm{u}$ is the flow velocity vector written in terms of the flow domain coordinate $\bm{x}\in \Omega$ ($\Omega$ denotes the flow domain). The body force $\bm{f}_b$ is implicitly computed at each coordinate on the airfoil, $\bm{\xi}\in\Gamma$ ($\Gamma$ represents the airfoil surface), to ensure the flow velocity matches the surface velocity $\bm{u}_b$. This surface velocity is prescribed to have normal component given by (\ref{eqn:normal_vel}) on the suction surface, and to have zero value in the tangent direction on the suction surface and zero value in both the tangent and normal directions on the pressure surface. In the governing equations (\ref{eqn:NS})--(\ref{eqn:BC}) all lengthscales are nondimensionalized by the chord length $c$, velocity scales by the freestream speed $U$, and time scales by the convective time $c/U$. 

The governing equations (\ref{eqn:NS})--(\ref{eqn:BC}) are spatially discretized using a discrete streamfunction-vorticity formulation using the standard second order finite volume stencils, removing the need to compute the pressure. The terms involving the Dirac delta function in (\ref{eqn:NS}) and (\ref{eqn:BC}) are approximated via a regularized delta function and discretized using a trapezoid-based quadrature scheme. The spatially discrete equations are discretized in time using the trapezoid method for the viscous term and the second-order Adams-Bashforth method for the advection term. The flow equations are treated using a multidomain approach: the finest grid surrounds the body and grids of increasing coarseness are used at progressively larger distances from the airfoil. See \citet{colonius2008fast} for more details. The simulation parameters and domain independence results are reported in appendix \ref{appA}.

\section{Surface morphing at an angle of attack of $\alpha = 5^\circ$}

We present in this article results for angles of attack of $\alpha=5^\circ$ and $\alpha=15^\circ$. The former angle of attack, considered in this section, involves a steady unactuated flow that offers a natural starting point for studying the spatio-temporal flow variations resulting from morphing.

In section \ref{sec:AOA15}, we consider the latter angle of attack and probe the interplay between the temporally and spatially varying morphing kinematics with an unactuated flow that has intrinsic vortex shedding behavior.

\subsection{Characteristics of the unactuated flow}
\begin{figure}[t]
    \centering
    \includegraphics[width =1.00\textwidth]{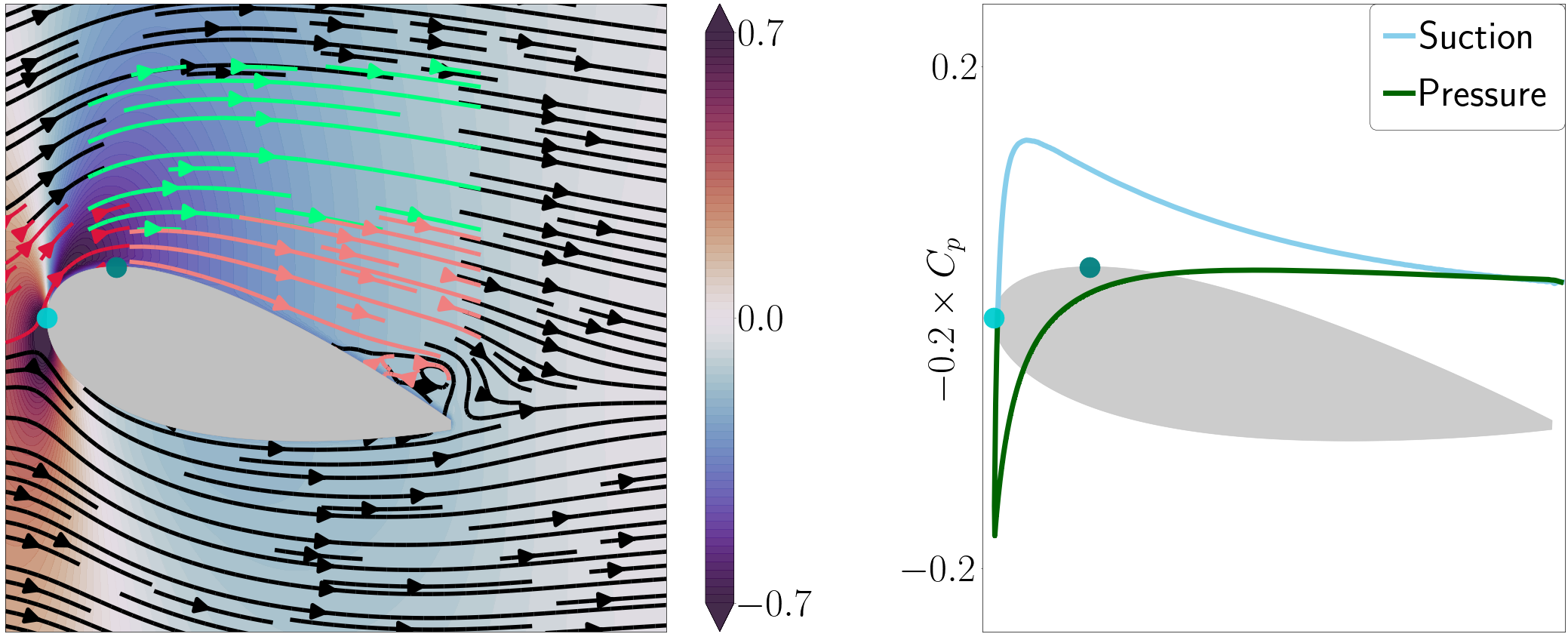}
    \caption{Features of the baseline (unmorphed) flow for the steady case of $\alpha=5^\circ$. Left: Coefficient of pressure, $C_p$, contour with streamlines. Right: $C_p$ distribution on airfoil surface (scaled by $-0.2$)}
    \label{fig:AOA5BLNew}
\end{figure}
To motivate the results involving actuation, we provide an overview of the flow field for the baseline/unactuated case (no morphing) at the angle of attack of $\alpha=5^\circ$. Figure %\ref{fig:AOA5BLP}
\ref{fig:AOA5BLNew} shows contours of the coefficient of pressure, $C_p$, with superimposed streamlines (left subplot) and the $C_p$ distribution on the suction and pressure sides of the airfoil (right subplot). The markers on the airfoil represent the leading edge in turquoise (henceforth, $s_{LE}$) and the point with the largest $y$-coordinate value (assuming the origin $(x,y)=(0,0)$ is attached to the leading edge, where the $x$-coordinate is along the freestream and the $y$-coordinate is in the vertical direction) in dark green (henceforth, $s_{max}$). The streamlines around the airfoil are shown in four different colors. The red streamlines represent the streamlines close to the airfoil surface between $s_{LE}$ and $s_{max}$. These streamlines illustrate the flow turning around the leading edge onto the suction surface, and are highlighted because the morphing velocity between $s_{LE}$ and $s_{max}$ will be shown to affect the curvature of these streamlines (and thus the local pressure field) considerably. The green streamlines indicate the streamlines downstream of $s_{max}$, above the shear layer that forms when the flow reaches $s_{max}$ and encounters the adverse pressure gradient. It will be shown in subsequent sections that morphing near $s_{max}$ changes the curvature of these streamlines above the shear layer, and that there is an associated pressure signature. 
The dynamics of this pressure signature will be shown to be of prime importance in dictating aerodynamic performance. The streamlines in salmon fall between those in green and the airfoil surface and  represent vortical structures that arise close to the airfoil surface. All other streamlines are shown in black. The specific color delineations were chosen purely for graphical purposes, to illustrate the distinct physical phenomena that will be categorized in the presence of morphing.
In the unactuated case, there is a pressure peak on the suction surface between  $s_{LE}$ and $s_{max}$. This peak is spatially correlated with the location where the red streamlines curve in the left figure. 
Beyond the point of minimum ${C_p}$, there is pressure recovery and the streamlines are only slightly convex. There is also a small region of separated flow near the trailing edge that has a minimal impact on the local pressure. 

\subsection{Surface morphing: lift dynamics and performance maps}
\begin{figure}
  \centering
    \subcaptionbox{\label{fig:fmAOA5_I}}{\includegraphics[width=3in]{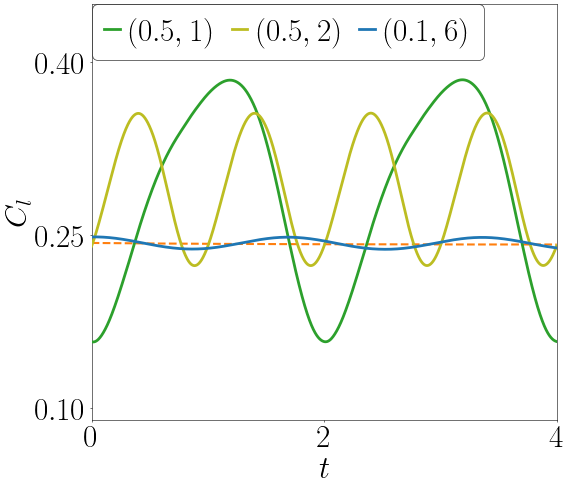}} 
    \hspace{0.4 cm}%
    \subcaptionbox{\label{fig:fmAOA5_II}}{\includegraphics[width=3in, height=2.65in]{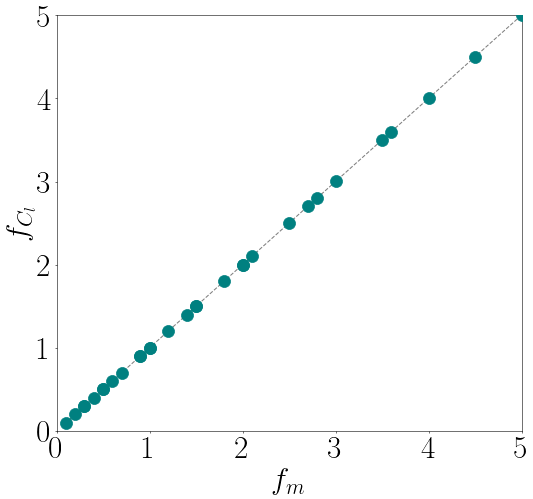}}
    \caption{Left: Temporal variation of $C_l$ for various wavespeed-wavenumber combinations, (\protect\baseline): baseline (unactuated) case. Legend represents $(w,n)$ values. Right: Peak frequency from a power spectral density analysis of temporal $C_l$ variation, versus $f_m$.}
    \label{fig:AOA5fm}
\end{figure}

In the presence of surface morphing, the lift dynamics are periodic, as shown for various wavespeed-wavenumber combinations, $(w,n)$, in figure \ref{fig:AOA5fm}. 
Figure \ref{fig:AOA5fm} also provides the peak frequency extracted from a power spectral density analysis of the lift signals for a variety of wavespeed-wavenumber combinations. For all morphing parameters considered at this angle of attack, the lift dynamics evolve at the frequency $f_m$. This fact demonstrates that in the presence of a steady unactuated flow, the dynamics are entirely driven by and evolve in concert with morphing.

\begin{figure}
    \centering
     \includegraphics[width =0.95\textwidth]{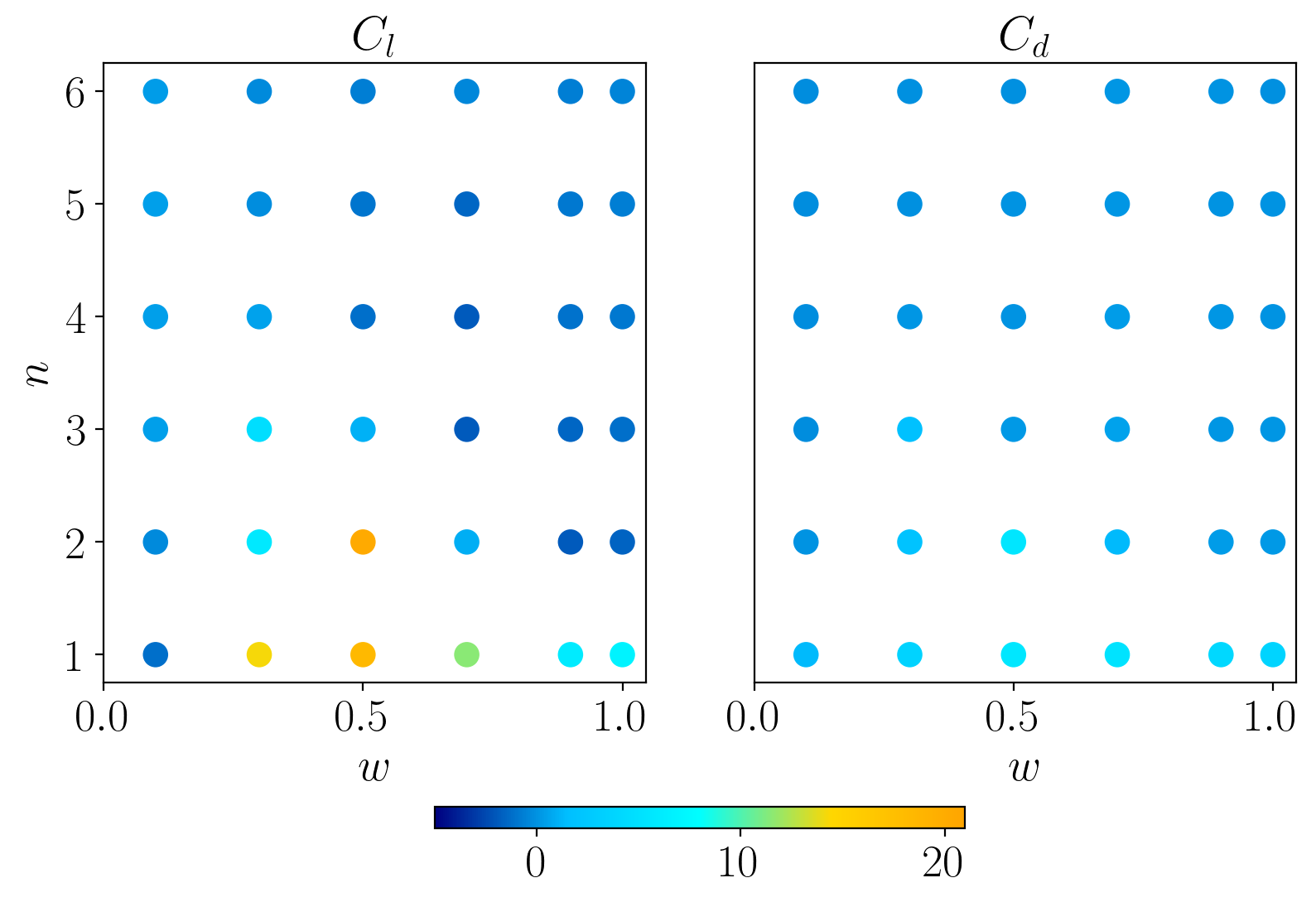}
    \caption{Performance Maps at $\alpha= 5^{\circ}$ showing the percentage increase in mean lift (left) and mean drag (right) as compared to the baseline (unactuated) case for different wavespeeds, $w$, and wavenumbers, $n$.}
    \label{fig:AOA5PM}
\end{figure}

The percentage change (relative to the baseline case) in coefficients of lift, $C_l$, and drag, $C_d$, are provided for various $w$ and $n$ values in figure \ref{fig:AOA5PM}.
Lift improvements are only observed for wavenumbers $n= 1$ and $2$. For both of these wavenumbers, maximum lift is achieved at $w=0.5$, with the global maximum occurring at a wavenumber of $n=2$. 

For higher values of $n$, the mean lift changes negligibly from the baseline (unactuated) case. The drag performance map demonstrates that all parameters lead to a drag penalty, though for the $(w,n)$ combinations leading to lift benefits the drag increases are significantly smaller than the lift improvements. In what follows, we therefore focus primarily on characterizing the effect that surface morphing has on the lift dynamics. 

\subsection{The effect of surface morphing on a motivating low-wavenumber example}
\begin{figure}[t]
    \centering
    \includegraphics[width =0.75\textwidth]{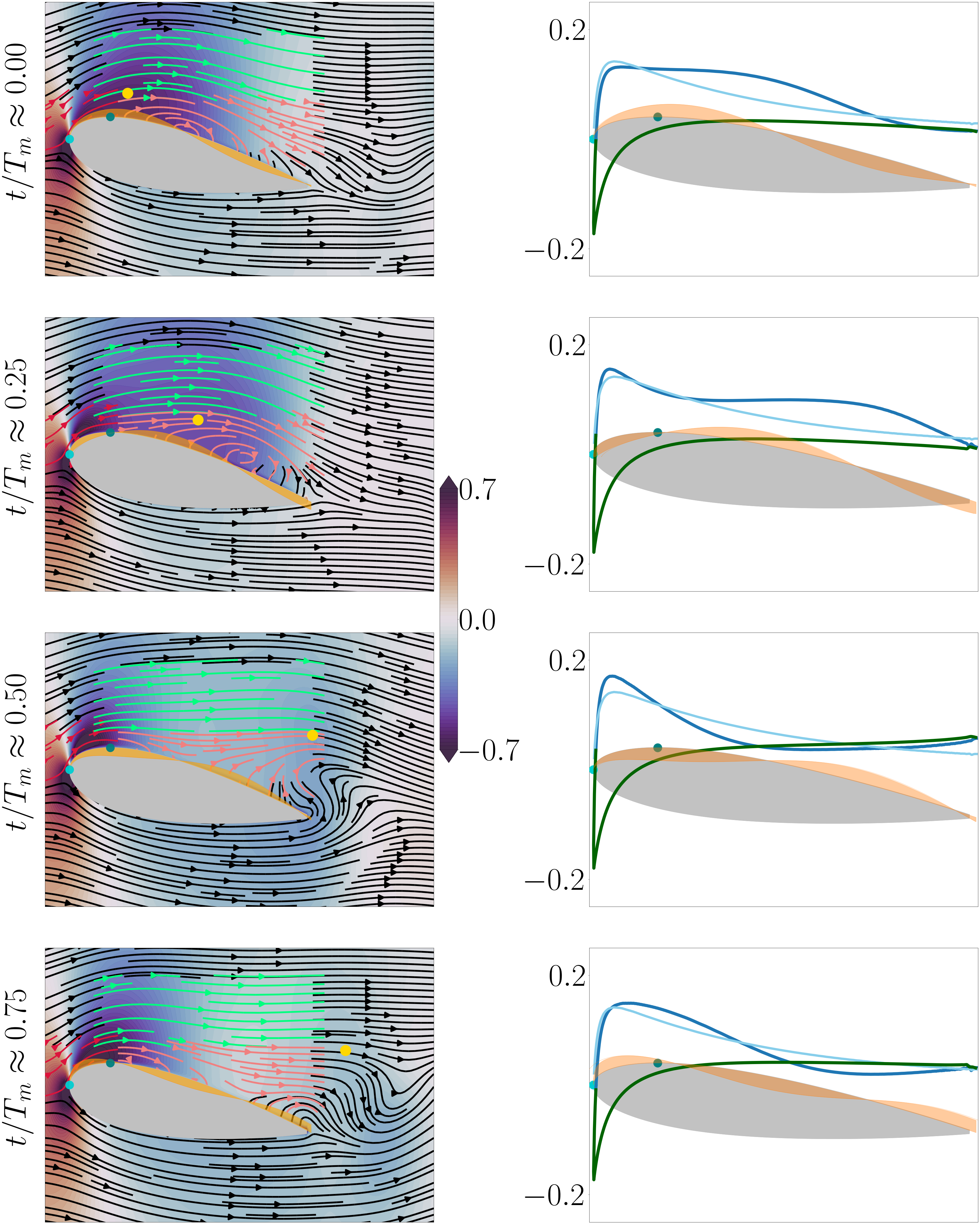}
        \label{fig:n1c0.3PI}
    \caption{$\alpha = 5^{\circ}$, $(w,n) = (0.3,1)$. Spatio-temporal variation of pressure due to morphing. Left:Coefficient of pressure, $C_p$, contour and streamlines. Right: $C_p$ distribution on airfoil surface (scaled by $-0.2$); dark blue with morphing on suction side, light blue without morphing on suction side; green with morphing on pressure side.}
    \label{fig:n1c0.3}
\end{figure}

We next probe the effect of morphing on the instantaneous flow field and pressure distribution. To motivate the nature of the analysis, we begin by considering the specific wavespeed-wavenumber combination $(0.3,1)$. This value was chosen because it lies within the parameter range where surface morphing has an appreciable effect on the mean lift. The physical processes identified for this parameter set will be shown in subsequent sections to persist for other parameters that are more beneficial to the mean lift. 

\begin{figure}
\centering
      \includegraphics[width =0.9\textwidth]{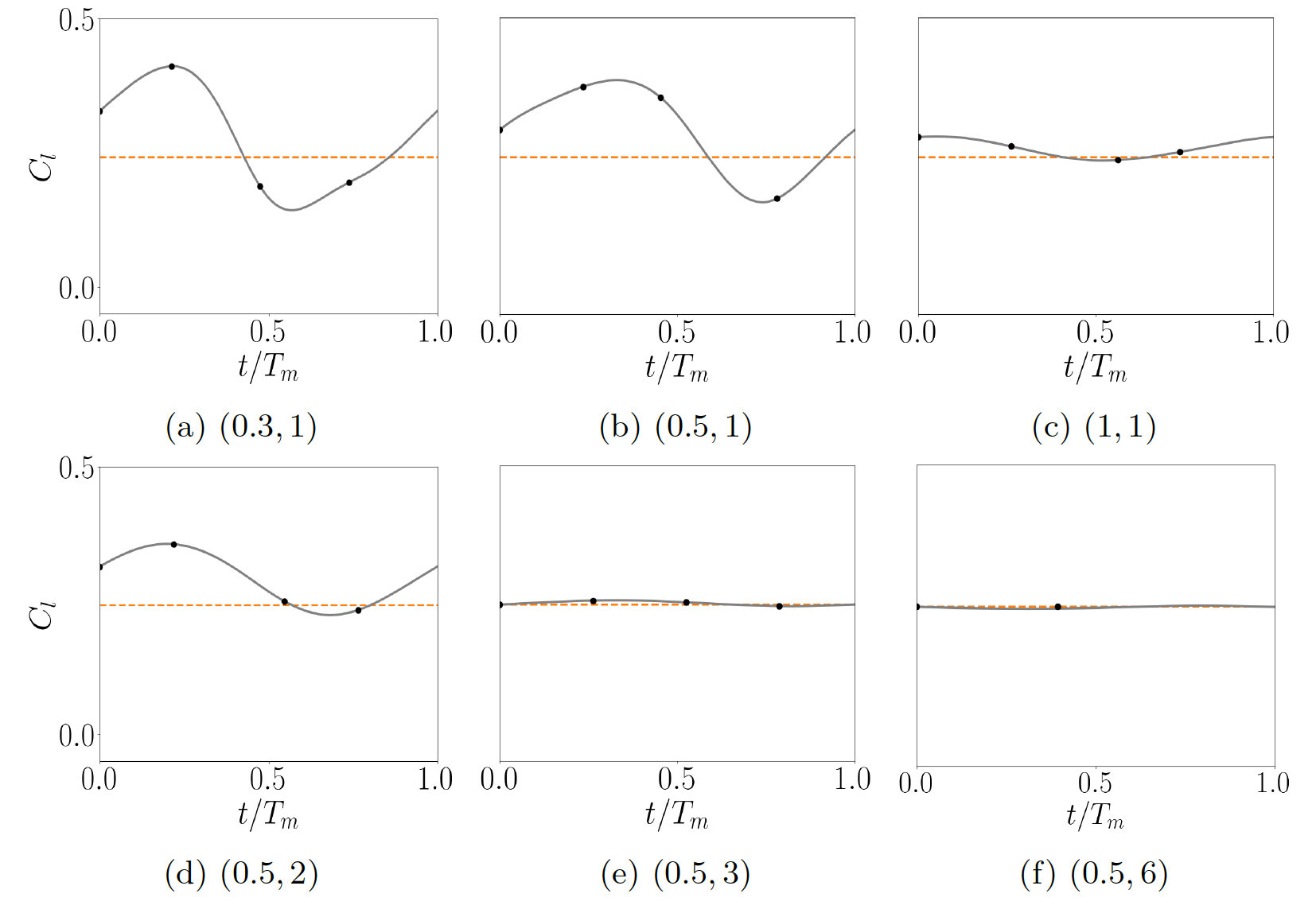}
      \caption{$\alpha = 5^{\circ}$. Temporal variation of $C_l$ as a function of morphing period for various $(w,n)$ combinations, (\protect\baseline) is the baseline value, (\protect\greyline) is the lift variation for the actuated flow. Markers represent instantaneous $C_l$ corresponding to snapshots in figures \ref{fig:n1c0.3}, \ref{fig:AOA5_n1_w0.5}--\ref{fig:n6_w0.5} respectively.}
    \label{fig:AOA5ClPlots}
\end{figure}

The left column of figure \ref{fig:n1c0.3} shows contours of coefficient of pressure, $C_p$, with superimposed streamlines (left subplot) (an exaggerated illustration of the morphing velocity is also shown for clarity) at four instances during the morphing cycle. The right column provides the associated $C_p$ distribution
along the airfoil. The temporal variation of coefficient of lift over one morphing period is shown in figure \ref{fig:AOA5ClPlots}a.

We first discuss the effect of surface morphing near the leading edge (between $s_{LE}$ and $s_{max}$), and subsequently describe the implications of this actuation further downstream along the airfoil. We will also make connections between the instantaneous pressure distribution and lift to explain why this wavespeed-wavenumber combination is suboptimal. At the time instance $t/T_m\approx0.5$, the morphing velocity between $s_{LE}$ and $s_{max}$ is negative and the stagnation point is below the leading edge. Commensurate with this, the red streamlines exhibit greater curvature than in the baseline case (c.f., figure \ref{fig:AOA5BLNew}). This behavior is associated with a larger pressure suction peak---relative to the baseline case---at the leading edge (right figure, third row). At the next time instance, $t/T_m\approx0.75$, the morphing velocity becomes positive between $s_{LE}$ and $s_{max}$. This change in velocity produces less pronounced curvature in the red streamlines, and an associated lessening of the pressure suction peak. This decrease in pressure suction across $s\in[s_{LE},s_{max}]$ continues when the morphing velocity becomes more positive in this region of the airfoil at the first time instance, $t/T_m\approx0.0$ (note that $t/T_m \approx 1$ is the same as $t/T_m \approx 0$). The trend finally reverses at $t/T_m\approx 0.25$, when the morphing becomes negative again and leads to an increase in the suction peak relative to the baseline case. These trends demonstrate that the pressure distribution at the airfoil along the leading edge is directly correlated with the surface morphing velocity over this region.

The effect of surface morphing over this region has implications on the pressure distribution  downstream of $s_{max}$ as well. The snapshots in the left column demonstrate that during instances when the the morphing velocity between $s_{LE}$ and $s_{max}$ is positive
($t/T_m\approx 0.75, 0$), the ``push-up'' effect from the positive velocity in that region, combined with the downward facing salmon streamlines downstream of $s_{max}$ (due to the negative morphing velocity there), 
lead to clockwise salmon streamlines downstream of $s_{max}$. Commensurate with this behavior is a greater curvature of the green streamlines above and near $s_{max}$, and thus a larger pressure suction peak, relative to the baseline case. This suction peak, which is indicated in the pressure contour plots via a yellow marker has a signature in the pressure distribution along the airfoil as well: the airfoil pressure distribution has a local suction peak near the $x$-location of the yellow marker. 
The opposite effect is seen when the morphing velocity between $s_{LE}$ and $s_{max}$ is negative
($t/T_m\approx 0.25, 0.5$): the negative morphing velocity near the leading edge leads to streamlines that wrap more tightly around the airfoil as they move from $s_{LE}$ to $s_{max}$. 

The result of these effects is flatter green streamlines and a lack of pressure suction peak in the flow field or on the airfoil (apart from the peak near the leading edge that is always present, including in the baseline case, due to the increased flow velocity before the adverse pressure gradient is encountered).

One of the more conspicuous features of the yellow marker is that it advects at a speed that is distinct from the morphing wavespeed: the marker begins near the morphing velocity peak when the peak is near $s_{max}$, but outpaces the morphing velocity peak as the two move downstream. This fact suggests that the pressure suction signature initiated near $s_{max}$ by a positive morphing velocity between $s_{LE}$ and $s_{max}$ then moves downstream at a speed that is agnostic to the morphing velocity, and in a manner that affects the local suction peaks in the pressure distribution on the airfoil. This discrepancy between the advection speed and the wavespeed of surface morphing also suggests a reason why this parameter set is sub-optimal for lift performance: because the suction peak moves faster than the morphing velocity peak, there will be periods of time where the suction peak has moved downstream of the airfoil before the morphing velocity peak has made its way all the way downstream and back to the leading edge. 
This can be further elucidated by comparing lift at the different instances of the snapshots (c.f., figure \ref{fig:AOA5ClPlots}a). At time instances $t/T_m=0, 0.25$ when the curvature of the green streamlines grows and advects downstream, there is an increase in lift while at $t/T_m = 0.5, 0.75$ when the suction peak moves downstream of the airfoil, lift is lower.
These time instances %periods 
have associated absences of local lift peaks. One would thus expect that faster morphing speeds, commensurate with the advection speed of the pressure suction signature indicated by the yellow marker, would lead to a pressure suction peak that is present over the entire period, leading to more lift benefits. 

\begin{figure}
    \centering
        \includegraphics[width =0.6\textwidth]{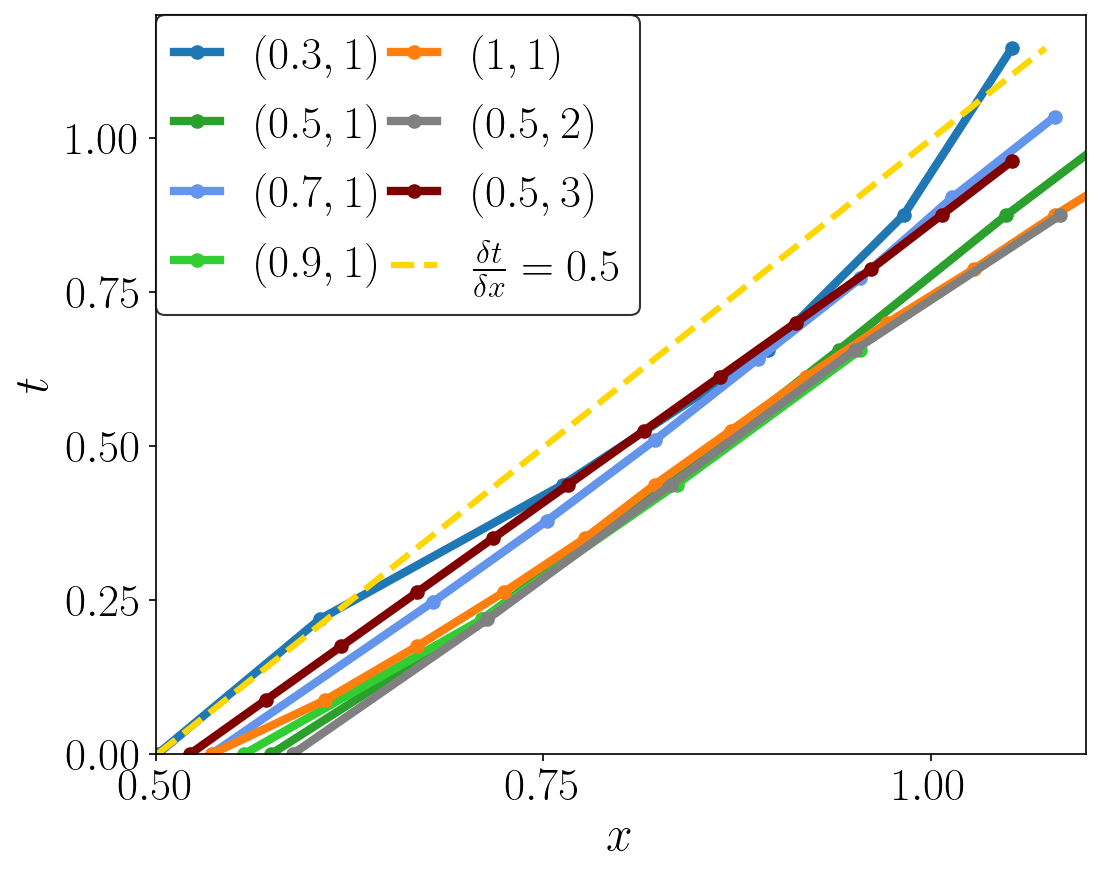}
    \caption{$\alpha = 5^{\circ}$, $t$-$x$ diagram of the local pressure minimum for various wavespeed-wavenumber combinations. Legend represents $(w,n)$ values.}
    \label{fig:AOA5_AdvVel}
\end{figure}

To demonstrate that the advection speed of this pressure suction peak is indeed agnostic to the morphing wavespeed, we show in figure \ref{fig:AOA5_AdvVel} a $t$-$x$ diagram of the local pressure minimum for various morphing parameters. Note that the time axis is in terms of convective time and is not scaled by the morphing period. The advection velocity is close to $0.5$ for all wavespeed-wavenumber combinations. Recalling the performance maps from figure \ref{fig:AOA5PM}, figure \ref{fig:AOA5_AdvVel} demonstrates that optimal performance occurs when the surface morphing kinematics are attuned to the intrinsic flow dynamics. 

Finally, we note that morphing leads to changes on the pressure side of the airfoil as well. In figure \ref{fig:n1c0.3}, as the local pressure minimum (indicated by the yellow marker) moves towards the trailing edge, there is a decrease in pressure on the pressure side near the trailing edge. This effect is most clearly seen by comparing the flow field at 
$t/T_m \approx 0.25$ and $t/T_m \approx 0.5$. The pressure near the trailing edge on the pressure side is lower at 
$t/T_m \approx 0.5$. This effect of morphing on the trailing edge will be shown to be correlated to the change in formation of the trailing edge vortex 
for the higher angle of attack case of $\alpha = 15^\circ$ in section \ref{sec:AOA15} below.

\subsection{Influence of surface morphing for low $n$ relevant to lift benefits}

\begin{figure}
    \centering
        \includegraphics[width =0.95\textwidth]{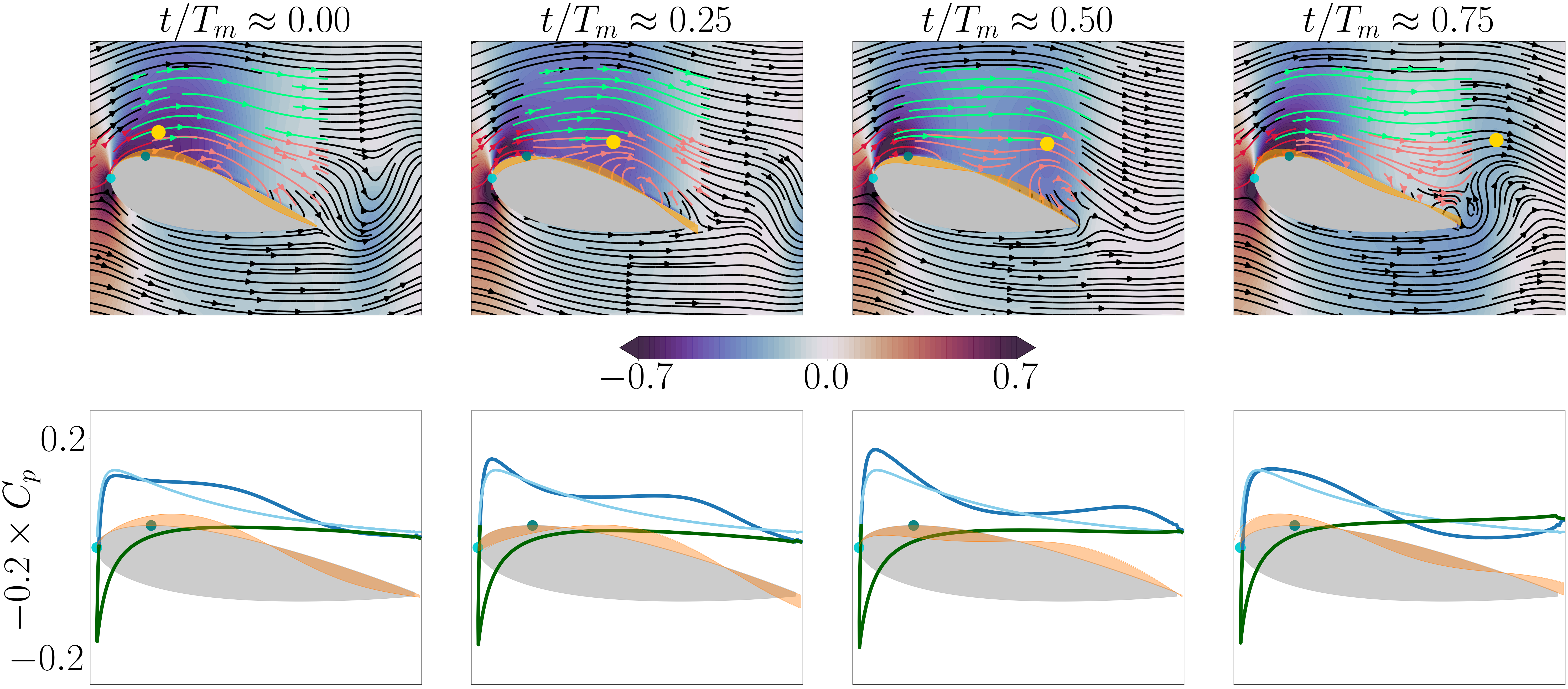}
        \label{fig:n1c0.5PI}
    \caption{Analog of  figure \ref{fig:n1c0.3}; $(w,n) = (0.5,1)$}
    \label{fig:AOA5_n1_w0.5}
\end{figure}

\begin{figure}
        \centering
        \includegraphics[width =0.95\textwidth]{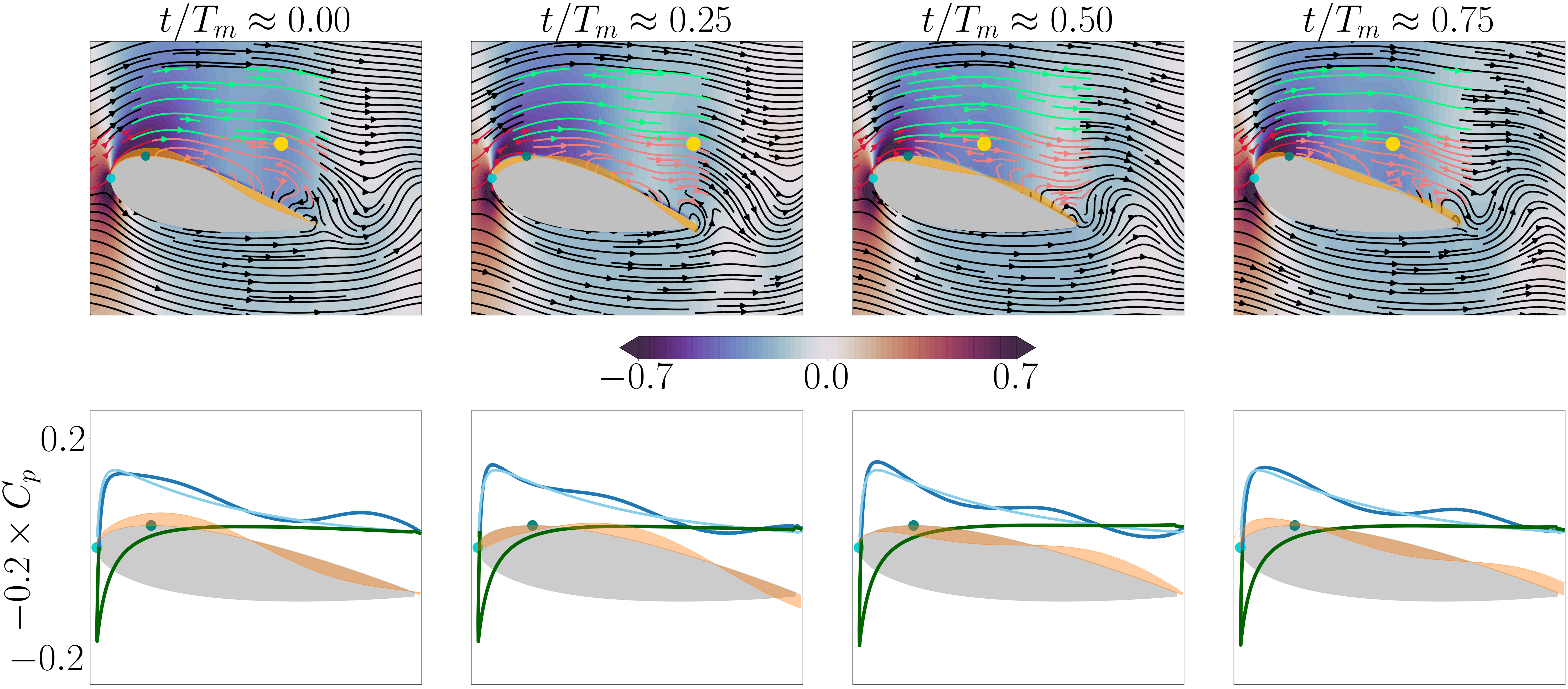}
        \label{fig:n1c1PI}
    \caption{Analog of  figure \ref{fig:n1c0.3}; $(w,n) = (1,1)$}
    \label{fig:AOA5_n1_w1}
\end{figure}

We now consider how the surface morphing wavespeed, $w$, and wavenumber, $n$, affect the system dynamics. We begin by fixing the wavenumber at $n=1$ to isolate the effect of wavespeed. Figures \ref{fig:AOA5_n1_w0.5} and \ref{fig:AOA5_n1_w1} provide analogous information to that of figure \ref{fig:n1c0.3}, for the parameter sets $(w,n)$ given by $(0.5,1)$ and $(1,1)$, respectively. The lift variation corresponding to these two wavespeed-wavenumber combinations are shown in figure \ref{fig:AOA5ClPlots}(b) and (c), respectively. 
A comparison of figures \ref{fig:n1c0.3}, \ref{fig:AOA5_n1_w0.5} and \ref{fig:AOA5_n1_w1} shows that many of the same features identified in the previous section persist: the pressure suction peak on the airfoil near the leading edge is correlated with the morphing velocity in this region. In addition, when the morphing velocity is positive near the leading edge, there is a pronounced streamline curvature near $s_{max}$  (illustrated via the green streamlines) and an associated local pressure minimum.
This pressure minimum, along with its signature via a local suction peak on the airfoil pressure distribution, advects downstream as time evolves. 
For the wavespeed $w=1$, there are two pressure peaks at time instances $t/T_m= 0, 0.25$. The snapshots in figure \ref{fig:AOA5_n1_w1} only highlight the second peak. However at $t/T_m=0.5, 0.75$,  the first peak from earlier instances is highlighted.

The figures also bear out the result from figure \ref{fig:AOA5_AdvVel} that the advection of the pressure minimum in the flow field occurs at a distinct speed from the morphing wavespeed: the yellow marker moves downstream of the airfoil faster, roughly commensurate with, and slower than the morphing velocity peak for $w=0.3$, $0.5$, and $1$, respectively. This suggests a reason for optimal lift benefits when $w=0.5$ at a given (low) value of $n$: mean lift is maximized when the morphing dynamics are synchronized with the intrinsic timescales of the aerodynamic system.

The optimal mean lift for $w=0.5$ can also be interpreted in terms of the morphing period (since $n$ is fixed for these three cases). For smaller wavespeeds $w$, the morphing period is longer than the amount of time it takes the local pressure suction peak to advect downstream, leading to time instances where this lift-producing pressure reduction is absent. The reduction in lift below the baseline value seen in figure \ref{fig:AOA5ClPlots}(a) is reflective of instances when the local pressure minimum (and thereby lift-beneficial suction peak) has advected downstream of the airfoil.
By contrast, larger values of $w$ lead to at least one local suction peak at all time instances. This observation unto itself would suggest that larger values of $w$ are better for lift production. Yet, the performance benefits are maximal at the intermediate value of $w=0.5$. This fact is due to the decrease in the magnitude of the pressure suction peak if $w$ grows too large, which is reflected in the lower value of $C_l$ increase compared with the baseline (c.f., figure \ref{fig:AOA5ClPlots}(c) and the smaller magnitude of the pressure peaks from figure \ref{fig:AOA5_n1_w1}).


Indeed, the airfoil pressure suction peak (shown in the right column of figure  \ref{fig:n1c0.3} and bottom row of figures \ref{fig:AOA5_n1_w0.5}, \ref{fig:AOA5_n1_w1})---that advects downstream along with the pressure minimum indicated by the yellow marker in the corresponding $C-p$ contours---
%top row of the subfigures---
is comparable in magnitude for $w=0.3$ and $w=0.5$ and smaller for $w=1$.
These observations suggest that lift benefits are obtained by a balance between (i) having a sufficiently fast morphing wavespeed that a pressure suction peak is present at all time instances and (ii) having a sufficiently slow wavespeed that the pressure suction peak has the largest magnitude possible. This balance is struck when the timescales of morphing and the bulk flow dynamics above the shear layer align. 

Noting from figure \ref{fig:AOA5PM} that maximum mean lift occurs for wavespeed-wavenumber values of $(w,n)=(0.5,2)$, we focus now on the effect of wavenumber and consider wavenumbers of $n=1$, $2$, $3$ for a fixed $w=0.5$. The lift dynamics in
figure \ref{fig:AOA5ClPlots} demonstrate that the optimal wavenumber for mean lift, $n=2$, produces a lower lift maximum than for $n=1$, but also has a higher lift minimum. This fact demonstrates that optimal lift for this parameter set is obtained through a balance between maintaining sufficiently high (but not optimal) lift maxima while mitigating the lift reducing troughs. 

\begin{figure}
    \centering
        \includegraphics[width =0.95\textwidth]{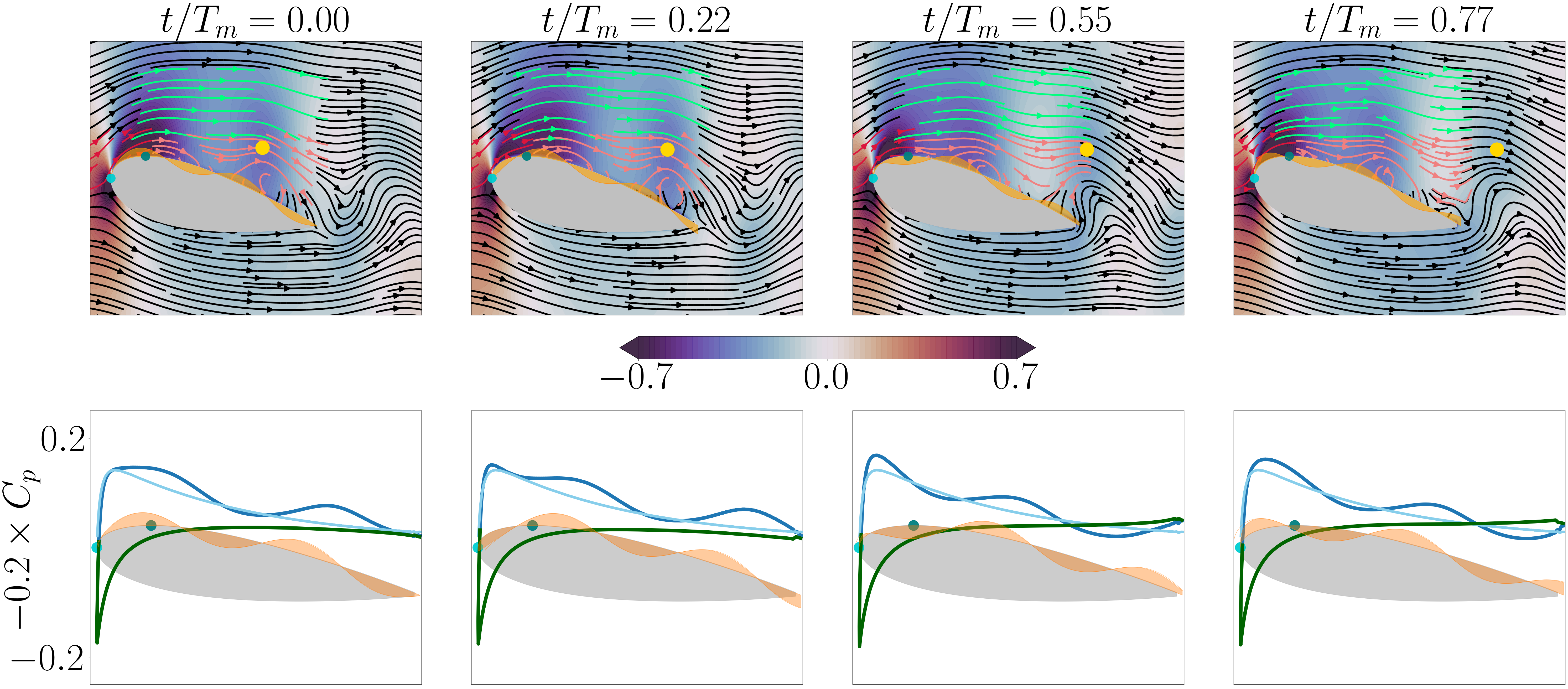}
        \label{fig:n1c1PI}
    \caption{Analog of figure \ref{fig:n1c0.3}; $(w,n) = (0.5,2)$}
    \label{fig:n2w0.5}
\end{figure}
\begin{figure}
        \centering
        \includegraphics[width =0.95\textwidth]{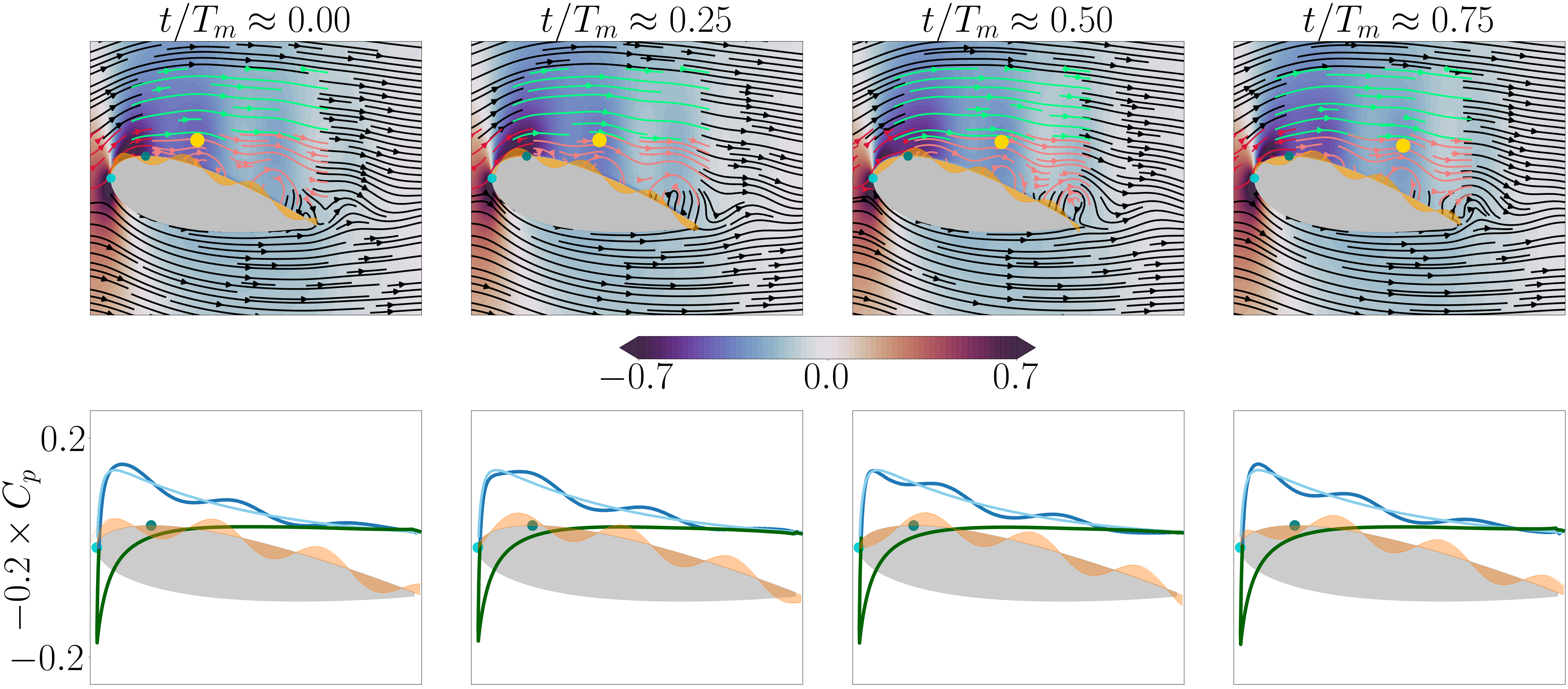}
        \label{fig:n1c1PI}
    \caption{Analog of figure \ref{fig:n1c0.3}; $(w,n) = (0.5,3)$}
    \label{fig:n3w0.5}
\end{figure}

To clarify the mechanisms by which this balance is obtained, we provide in figures \ref{fig:n2w0.5} and \ref{fig:n3w0.5} analogous plots to those of figure \ref{fig:AOA5_n1_w0.5}, for the wavespeed-wavenumber values of $(w,n)=(0.5,2)$ and $(0.5,3)$. The 
bottom rows of figures \ref{fig:AOA5_n1_w0.5}, \ref{fig:n2w0.5}, and \ref{fig:n3w0.5} demonstrate that at time instances near the maximum lift, $t/T_m\approx 0.25$, the number of local suction peaks of the airfoil pressure distribution increases with the wavenumber $n$. This result is intuitive from the discussion above: as each peak in the morphing velocity moves from $s_{LE}$ to $s_{max}$, a new local pressure minimum (indicated by the yellow markers) is created in the flow field near $s_{max}$ and there is an associated signature in the airfoil pressure distribution. This local pressure minimum and its corresponding local suction peak in the airfoil pressure distribution advect downstream at the bulk advection velocity of $\sim 0.5$ (c.f., figure \ref{fig:AOA5_AdvVel}). For higher wavenumber values, this process of generating these local pressure minima occurs multiple times before a given local pressure minimum advects downstream of the airfoil. This process of generating several local pressure peaks is beneficial towards avoiding low lift minima. Viewing the 
bottom rows
of figures \ref{fig:AOA5_n1_w0.5}, \ref{fig:n2w0.5}, and \ref{fig:n3w0.5} at time instances near the minimum lift, $t/T_m \approx 0.75$, it is clear that lift-deteriorating troughs in the airfoil pressure distribution, which appear between the local suction peaks, are lessened in magnitude. 

The ability to mitigate these lift minima, however, also leads to lower maximum lift. At the aforementioned time instances near maximum lift, figures \ref{fig:AOA5_n1_w0.5}, \ref{fig:n2w0.5}, and \ref{fig:n3w0.5} show that the local suction peak in the airfoil pressure distribution is largest for a wavenumber of $n=1$ and smallest for $n=3$. This decrease in magnitude of the local suction peaks of the airfoil pressure distribution can be intuited by observing that at the fixed wavespeed $w=0.5$, increasing the wavenumber increases the frequency $f_m=wn$ over which a segment of the wave will repeat itself in time. This increase in frequency creates a smaller time window over which morphing can impact the flow at a given location before there is a change in actuation direction. 

Altogether, these results indicate that increasing the wavenumber introduces competing effects of (i) the ability to create multiple local pressure minima along the airfoil surface to mitigate lift detriments and (ii) a decrease in the magnitude of these suction peaks, resulting in lower lift maxima. 

\subsection{Higher $n$ values: Insignificant effect of surface morphing}

\begin{figure}
    \centering
        \includegraphics[width =0.85\textwidth]{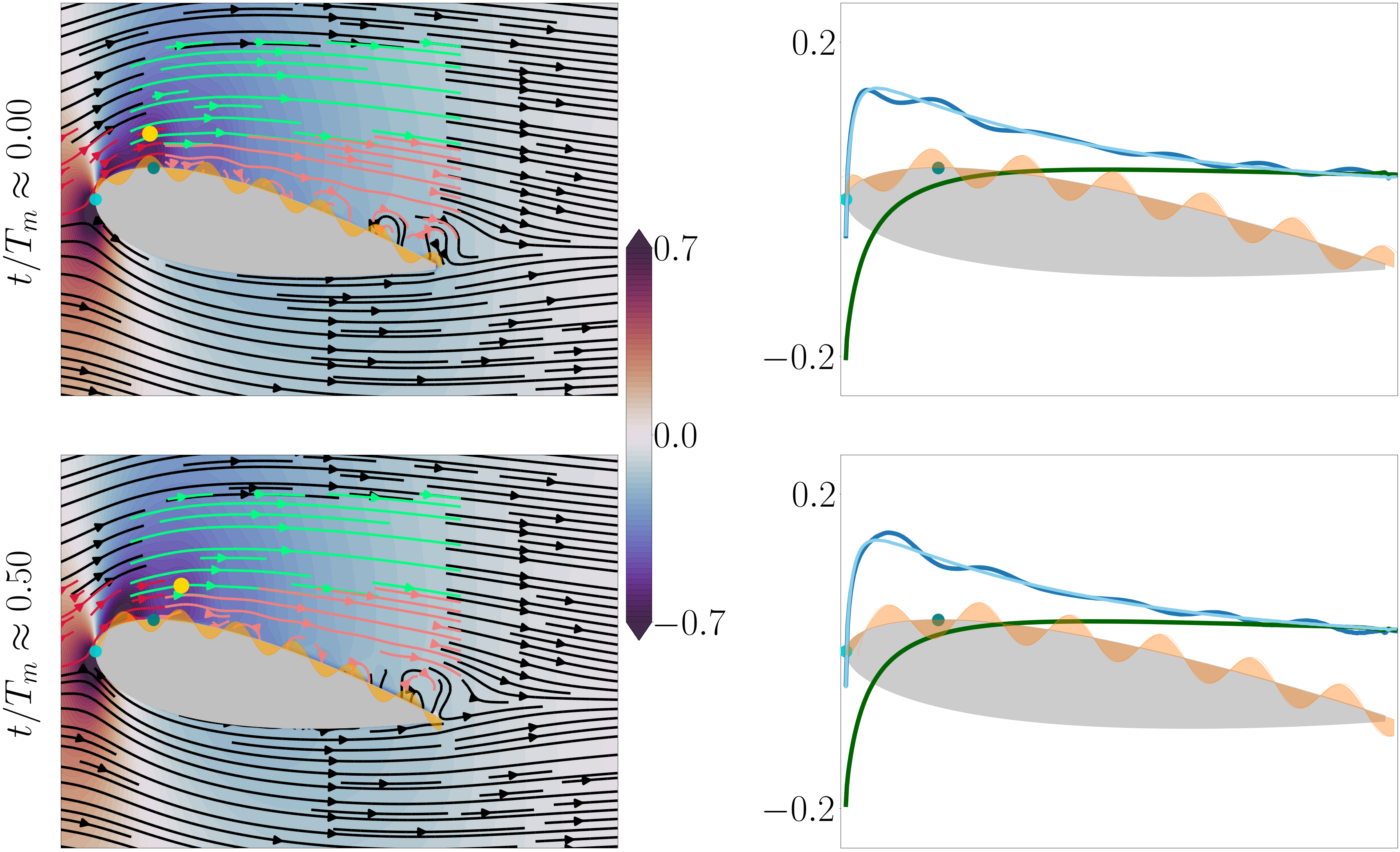}
        \label{fig:n6c0.5PI}
    \caption{Analog of figure \ref{fig:n1c0.3}; $(w,n) = (0.5,6)$}
    \label{fig:n6_w0.5}
\end{figure}

For higher wavenumber values ($n\ge 3$), surface morphing has a negligible impact on lift or drag (c.f., the performance maps in figure \ref{fig:AOA5PM} and the time traces of the lift dynamics in figures \ref{fig:AOA5fm} and \ref{fig:AOA5ClPlots}(f)). The decrease in effect of surface morphing can be understood through a timescale argument. For sufficiently large values of wavenumber $n$, the actuation frequency $f_m = wn$ will be large for all of the wavespeeds $w$ considered in this work. Commensurate with this increase in frequency is a reduction in the amount of time a given flow location is exposed to positive (negative) surface morphing before the actuation becomes negative (positive). For completeness, we provide in figure \ref{fig:n6_w0.5} analogous plots to those of figure \ref{fig:AOA5_n1_w0.5}, for the wavespeed-wavenumber values of $(w,n) \equiv (0.5,6)$.
The figure demonstrates that the green streamlines exhibit little curvature (and thus, insignificant low pressure regions in that area of the flow). There are instead a collection of circular streamlines near the airfoil surface where morphing occurs, though the small timescale of actuation at a given location reduces the extent to which surface morphing can impact the flow. This negligible change is visible in the plots of the airfoil pressure distribution as well. There are high-wavenumber fluctuations about the baseline (unactuated) pressure distribution, but with little net effect on lift.   
The high-wavenumber fluctuations in the airfoil pressure distribution can be correlated with the attributes of the streamlines close to the airfoil surface (shown in salmon). In the regions of attached flow, the clockwise streamlines are associated with reduced pressure. Near the trailing edge where the flow is reversed, the regions of reduced pressure coincide with anticlockwise streamlines. This is intuitive because clockwise streamlines along with flow towards the right would result in higher velocity. For reversed flow, anticlockwise streamlines would have the same effect.

The fact that 
the reduction in time leads to a negligible effect of actuation demonstrates that there are two separate timescales that the flow is receptive to, and others that the flow is not receptive to. The first timescale that is relevant for performance is the wavespeed, $w$. For sufficiently low wavenumber $n$, mean lift is maximized when the morphing wavespeed is aligned with the intrinsic advection speed above the shear layer. However, there is a second crucial timescale, the morphing frequency, that is encoded by the combined effect of wavespeed and wavenumber. To attain maximal lift benefits, this frequency must be within a range that the flow is receptive to. Said differently, for sufficiently high morphing frequencies the actuation has little impact on the flowfield, even when the %wavenumber 
wavespeed is attuned to the advection speed above the shear layer. 

\section{Surface morphing at an angle of attack of $\alpha= 15^{\circ}$}
\label{sec:AOA15}
 In this section, we consider a higher angle of attack, for which the unactuated flow exhibits unsteady separated-flow dynamics with prominent frequency signatures. We characterize the interplay between the morphing kinematics and the underlying vortex shedding behavior, leveraging insights from above about the key relationship between morphing and bulk advective processes in the flow as well as timescales over which the flow is receptive.

\subsection{Characteristics of the unactuated flow}
\begin{figure}[h]
    \centering
        \includegraphics[width =0.75\textwidth]{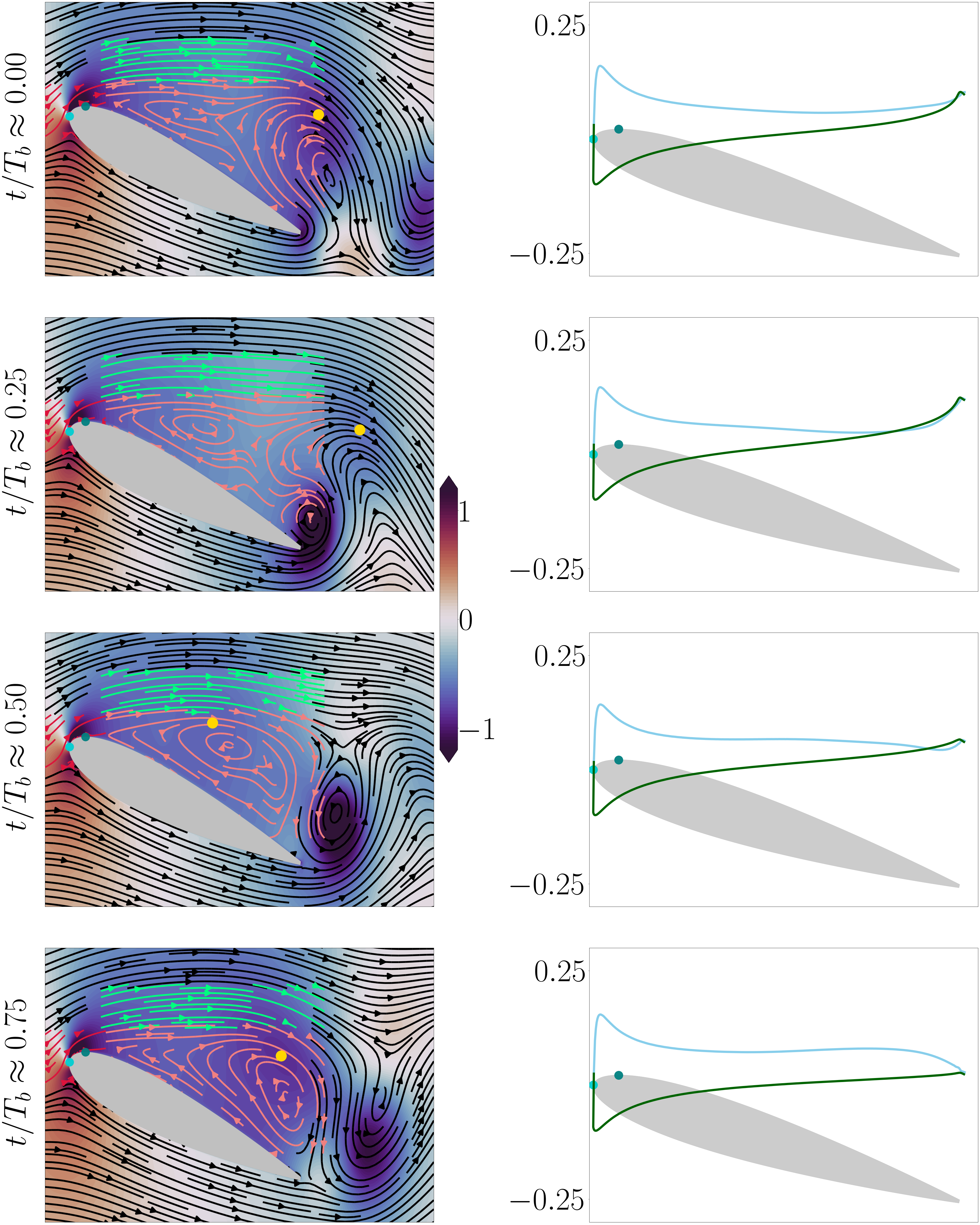}
    \caption{$\alpha = 15^{\circ}$, baseline flow. LEV and TEV formation. Left: Coefficient of pressure, $C_p$, contour and streamlines. Right: $C_p$ distribution on airfoil surface (scaled by $-0.1$), suction side in light blue, pressure side in green.}
    \label{fig:AOA15BL}
\end{figure}

At $\alpha=15^\circ$, the flow field in the absence of morphing is unsteady and periodic vortex shedding is seen. The frequency of shedding (henceforth, $f_b$) is $0.71$. The unsteadiness of the aerodynamic forces stems from the pressure variations on the suction and pressure sides of the airfoil, which correspond to the formation and advection of the leading- and trailing-edge vortices (LEV and TEV, respectively). Figure \ref{fig:AOA15BL} shows on the left contours of coefficient of pressure, $C_p$, in the flow field with superimposed streamlines, and on the right the $C_p$ distribution on the airfoil surface, at four instances of the shedding cycle. (The lift variation over one shedding cycle and the associated power spectral density of the lift dynamics are shown in figure \ref{fig:AOA15Regimes_FFT_New}a).

Over a lift period, the LEV and TEV successively form, advect, and interact. This process is highlighted through the flow-field snapshots on the left column. At $t/T_b \approx 0$, the LEV from the previous cycle is close to the trailing edge (and is strongly correlated with the local pressure minimum indicated by the yellow marker). Before the new LEV forms completely, the trailing edge vortex rolls up at $t/T_b \approx 0.25$. This instance where the new LEV has yet to roll up and the previous LEV has advected beyond the airfoil (there is no discernible pressure minimum along the airfoil) is associated with the minimal lift value; c.f., figure \ref{fig:AOA15Regimes_FFT_New}a. The minimum lift value is borne out in the $C_p$ distribution on the airfoil (right column, figure \ref{fig:AOA15BL}) via the evident pressure reduction on the pressure side. As the TEV advects downstream at $t/T_b \approx 0.5$, roll-up of the new LEV occurs and introduces a local pressure minimum (represented by the yellow marker). Commensurate with this pressure reduction is an increased curvature in the nearby green streamlines.
As the new LEV grows and advects from $t/T \approx 0.5$ to $t/T \approx 0.75$, the curvature of the green streamlines becomes more pronounced and there is greater pressure reduction. Associated with this behavior is maximal lift (c.f., figure \ref{fig:AOA15Regimes_FFT_New}a), manifesting in a lift-producing bump on the $C_p$ distribution on the airfoil suction side (c.f., the corresponding snapshot in the right column of figure \ref{fig:AOA15BL}). Note also the decrease in pressure magnitude %distribution 
on the pressure side compared with the minimal lift instance of $t/T_b \approx 0.25$.  This process in the unactuated case mirrors many features of the lower angle of attack scenario in the presence of surface morphing: both cases are driven by the appearance of flow structures near $s_{max}$, as well as their subsequent advection and manipulation of nearby streamlines and airfoil pressure signatures. This connection highlights an intrinsic propensity of this flow system. In what follows we will characterize how actuation further modulates these driving aerodynamic processes.

\subsection{Effect of morphing: general comments and regime characterization}

\begin{figure}
  \centering
    % \includegraphics[width =0.92\textwidth, height=0.75\textwidth]{AOA15_RegimeI.png}
    % \hspace{0.5em}\includegraphics[width =0.91\textwidth, height=0.53\textwidth]{AOA15_RegimeII.png}
    
    \includegraphics[width =0.92\textwidth,height=0.5\textwidth ]{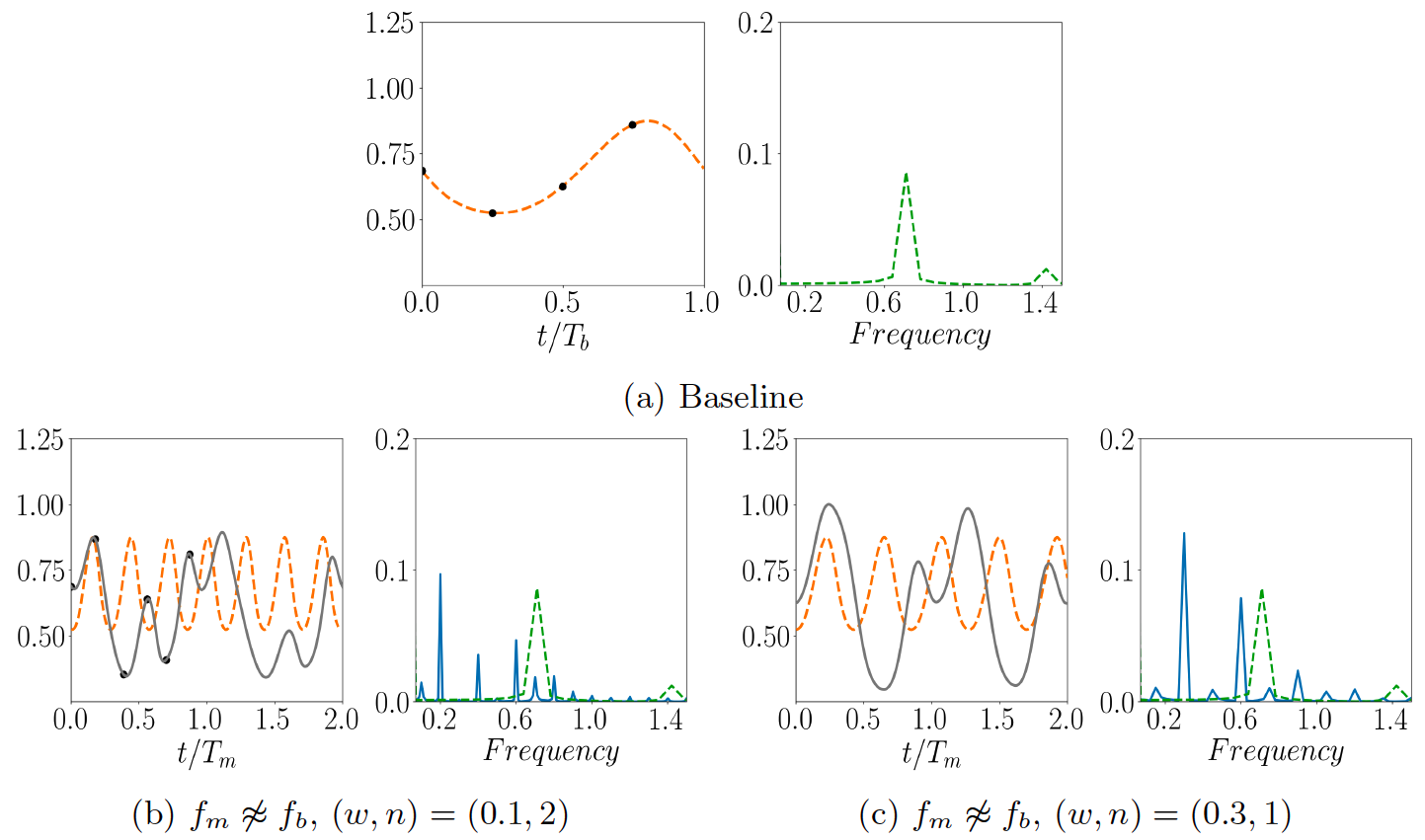}
    
    \vspace{0.5em}
    
    \hspace{0.15em} \includegraphics[width =0.92\textwidth,height=0.25\textwidth]{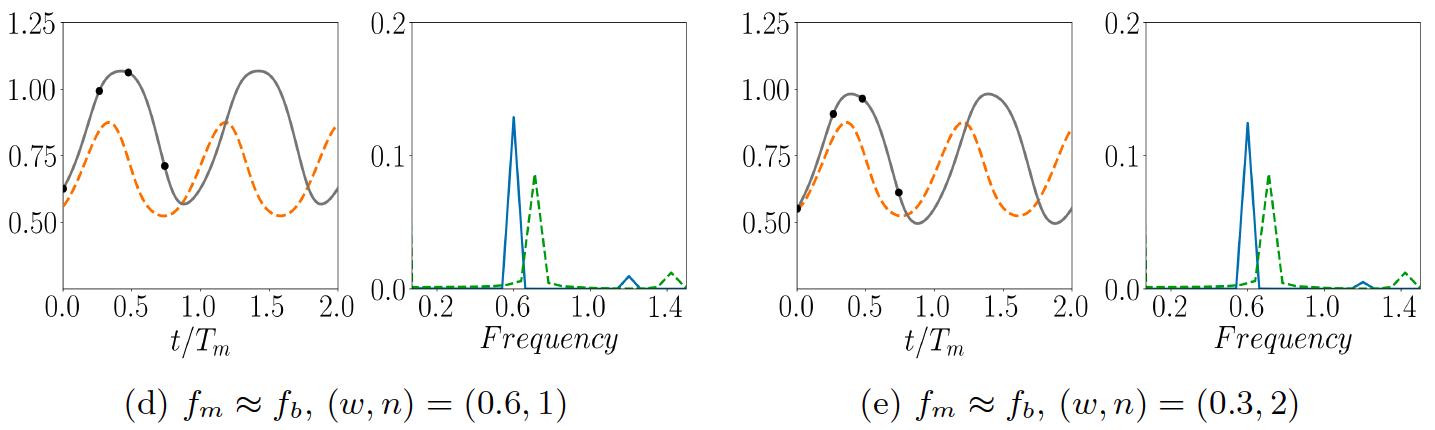}
    
    \vspace{0.5em}
    
     \hspace{0.1em} \includegraphics[width =0.92\textwidth,height=0.25\textwidth]{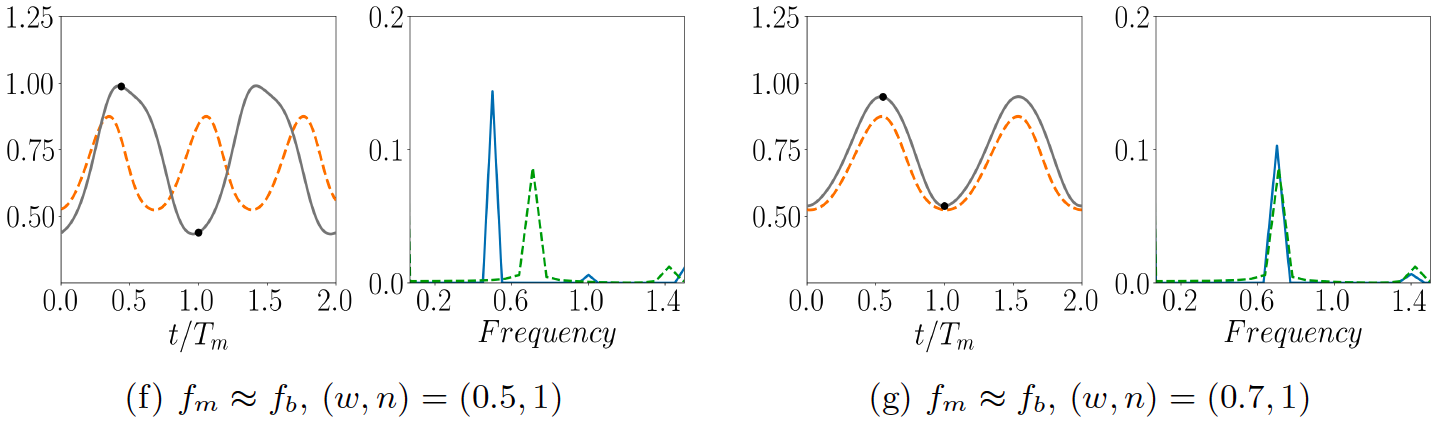}
    
    \vspace{0.5em}
    
     \hspace{0.35em} \includegraphics[width =0.92\textwidth,height=0.25\textwidth]{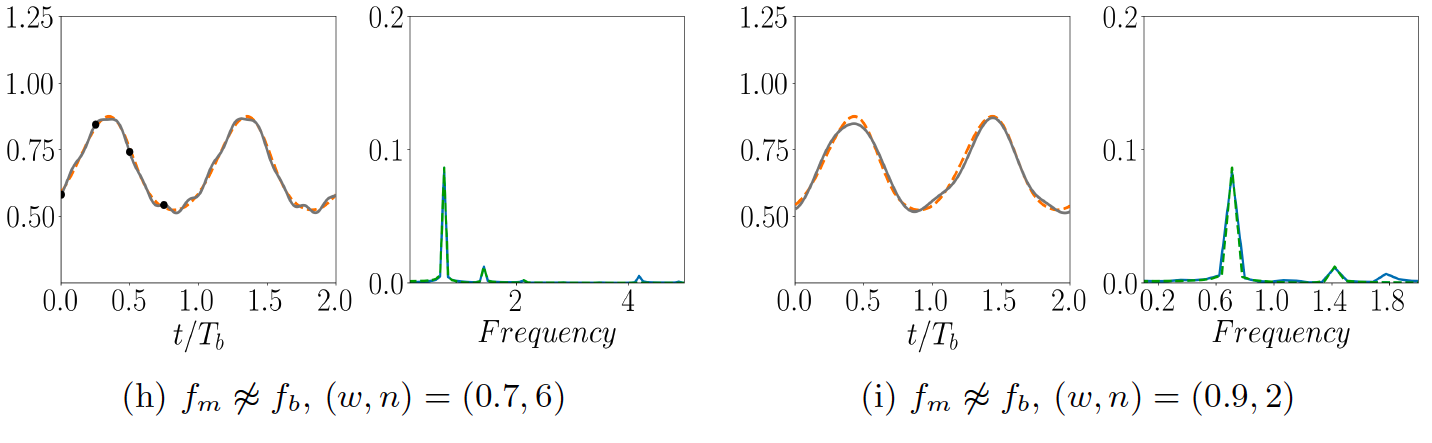}
    
    \caption{$\alpha = 15^{\circ}$. For each subfigure, left: temporal variation of $C_l$ for the baseline (\protect\baseline) and actuated (\protect\greyline) flows (markers correspond to instances at which snapshots are plotted in subsequent figures);  right: power spectrum of lift fluctuations for the baseline  (\protect\greenline) and actuated (\protect\blueline) flows.}
    \label{fig:AOA15Regimes_FFT_New}
\end{figure}

Whereas the actuated flow dynamics at the lower angle of attack evolved commensurately with the morphing timescales (c.f., figure \ref{fig:AOA5fm}b), figure \ref{fig:AOA15Regimes_FFT_New} shows three noticeably different classes of lift dynamics, which can be categorized primarily by the relationship between the frequencies of the lift curve in the baseline case, $f_b$, and the morphing frequency $f_m=wn$. We show in figure \ref{fig:AOA15Regimes_FFT_New} the lift dynamics and associated power spectral density (PSD) for parameters yielding these three separate regimes.

% \begin{figure}
%   \centering
%     \includegraphics[width =0.92\textwidth, height = 1.3\textwidth]{AOA15_Regimes.png}
%     \caption{$\alpha = 15^{\circ}$. For each subfigure, left: temporal variation of $C_l$ for the baseline (\protect\baseline) and actuated (\protect\greyline) flows (markers correspond to instances at which snapshots are plotted in subsequent figures);  right: power spectrum of lift fluctuations for the baseline  (\protect\greenline) and actuated (\protect\blueline) flows.}
%     \label{fig:AOA15Regimes_FFT_New}
% \end{figure}

 When $f_m \approx f_b$ (figures \ref{fig:AOA15Regimes_FFT_New}d--g), dynamics of the actuated flow can evolve along the morphing frequency (with small signatures at harmonic integers), with no discernible remnant of the unactuated frequency peaks. % remaining. 
 We will demonstrate below that over this regime, the flow is receptive to actuation and an analogous shedding process ensues as in the unactuated case, but over a timescale that is governed by the morphing frequency.  We therefore refer to this behavior as the lock-on regime. The potential for lock-on suggests that similar synchronization behavior can occur at low Reynolds numbers as was observed for high Reynolds numbers \citep{jones2018numerical,akbarzadeh2019numerical,akbarzadeh2019reducing}, though we will show that in this lower Reynolds number setting the result is not a reduction in flow separation. Instead, the vortex shedding processes remain but are modulated by morphing, in some cases to significant lift benefit. This phenomenon is a common occurrence in low-Reynolds-number aerodynamic flows; for example, the transient formation of the LEV in wings at low Reynolds numbers gives rise to temporary lift benefits \citep{Eldredge2019} and is a key mechanism behind insect flight \citep{dickinson1993unsteady}.

\begin{figure}
  \centering
    \includegraphics[width =0.92\textwidth]{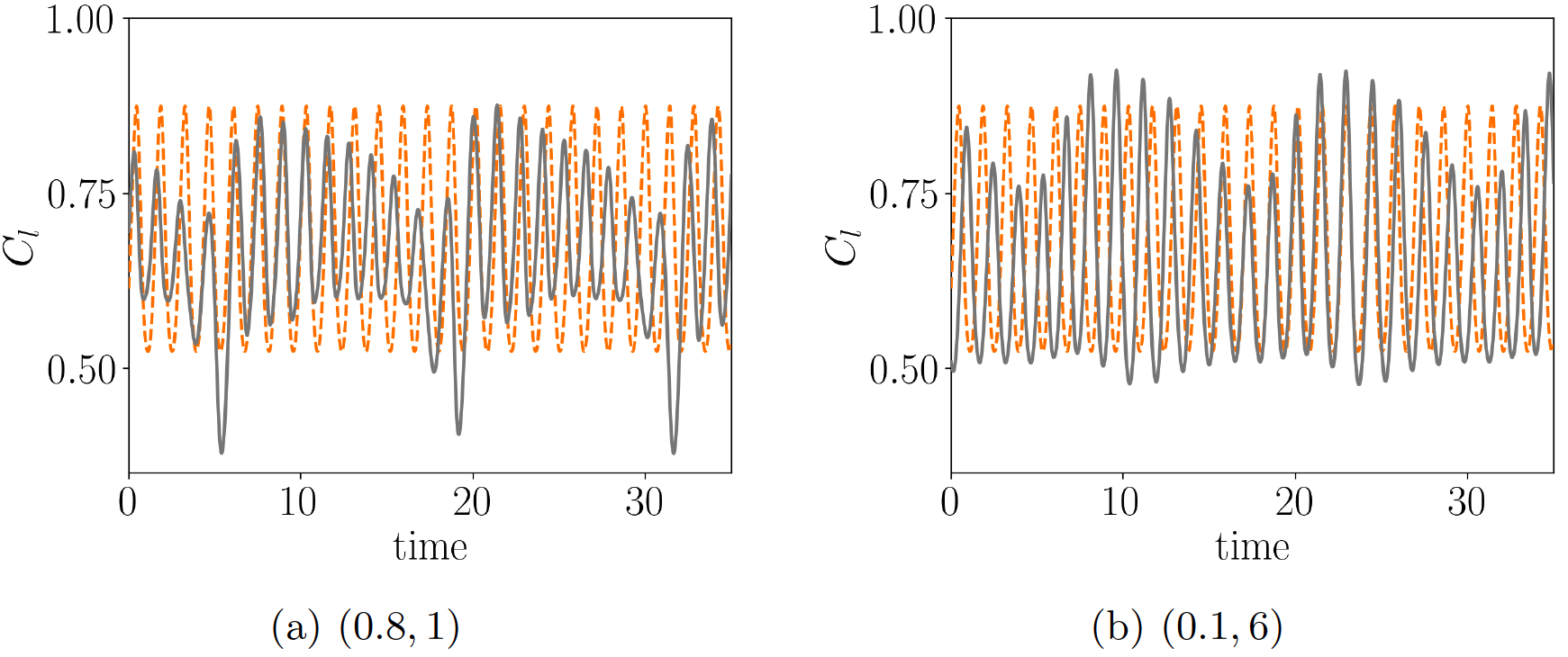}
    
    \includegraphics[width =0.92\textwidth]{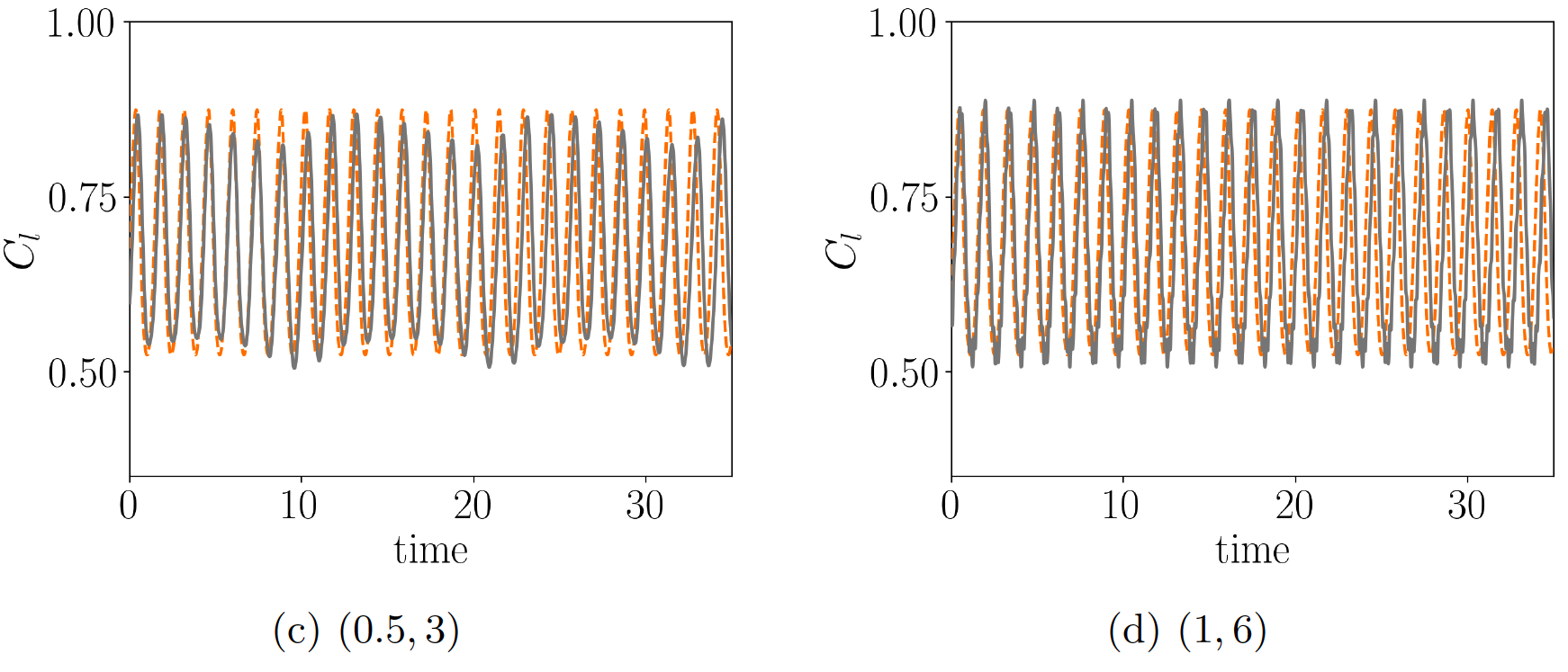}
    \caption{$\alpha = 15^{\circ}$. Temporal behavior of $C_l$ for cases selected from the interactive (a)-(c) and superposition (d) regimes. These cases occur when $f_m\not\approx f_b$ (i.e., when lock-on is not observed).}
    \label{fig:AOA15Liftbehaviour}
\end{figure}
\begin{figure}
    \centering
        \includegraphics[width =0.95\textwidth]{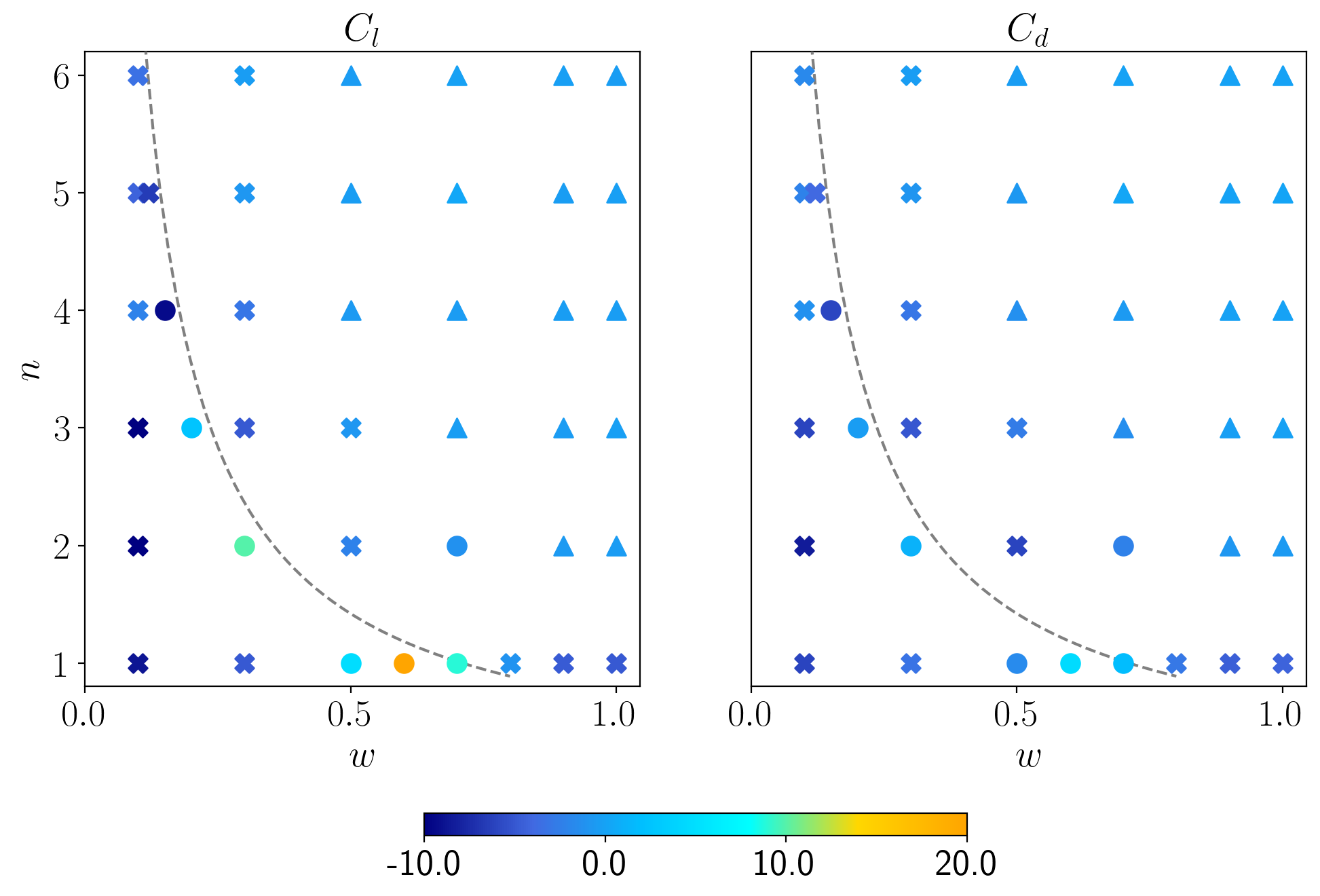}
    \caption{Performance Maps for $\alpha= 15^{\circ}$ showing the percentage change in mean lift (left) and mean drag (right) from the baseline (unactuated) case for different wavespeeds, $w$, and wavenumbers ,$n$. Circles, crosses, and triangles represent the lock-on, interactive and superposition regimes, respectively. The gray dashed line depicts $wn=0.71$ ($f_b$).  For the wavespeeds considered, lock-on is seen for $n \leqslant 4$. Note that the regime for  $(w,n) \equiv (0.7,2)$ (for which $f_m\approx 2f_b$) was marked as lock-on because  the dominant lift dynamics evolve at $f_b$ and morphing and lock-on were found to synchronize in the snapshots (not pictured). We refer to this case as harmonic lock-on. The existence of harmonic lock-on at higher wavenumbers was not found for the parameter values considered in this study, though it is possible that such behavior exists (although the regime map generally demonstrates that the propensity for lock on narrows as $n$ increases).}
    \label{fig:AOA15PM_NewII}
\end{figure}
\begin{figure}
    \centering
        \includegraphics[width =0.85\textwidth]{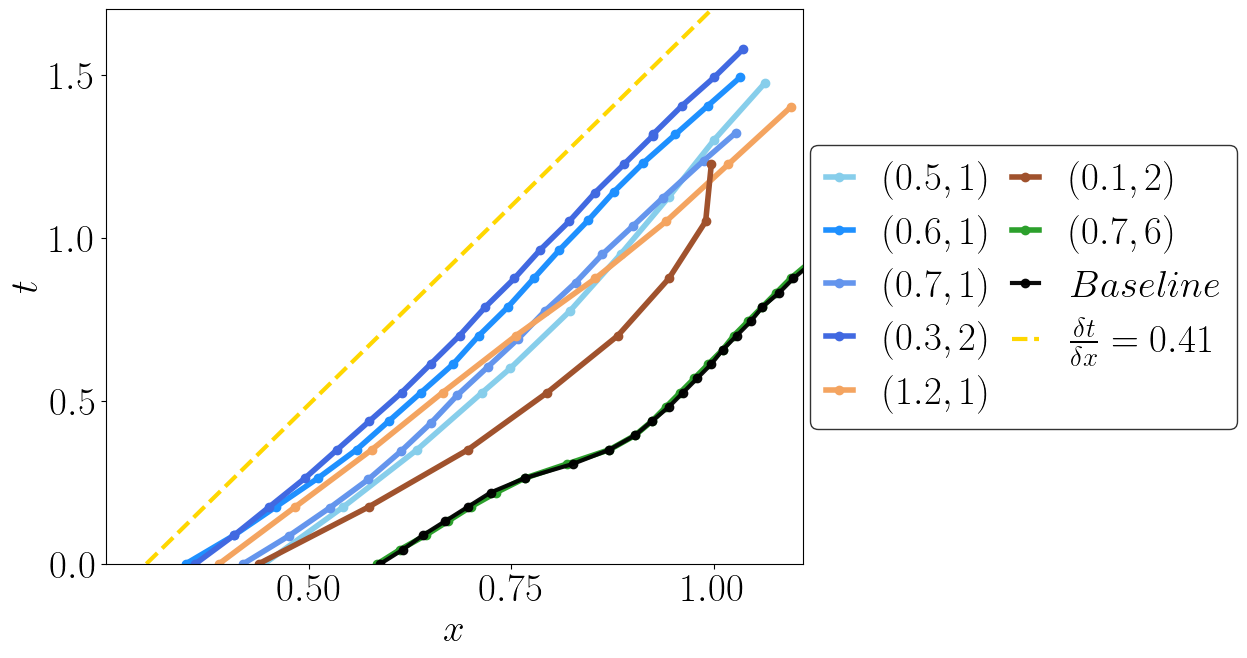}
    \caption{$\alpha = 15^{\circ}$, $t$ vs $x$ of the local pressure minimum near the airfoil (LEV) for the baseline (unactuated) case and morphing for various $(w,n)$ combinations. For $x$-locations without markers, there was no local minimum detected (coinciding in the subsequent snapshots to instances where the leading-edge vortex that produces the pressure minimum has not yet formed during the cycle).}
    \label{fig:LockOnAdvection}
\end{figure}
 When $f_m\not\approx f_b$, there are broadly two categories of behavior that can occur. In the first category, morphing and the underlying shedding behavior can interact in complex ways to produce a mixture of timescales. This behavior, termed here the interactive regime, is illustrated in figure \ref{fig:AOA15Regimes_FFT_New}b--c for cases where $f_m < f_b$: the time history of the lift curve conveys multi-scale dynamics that correspond in the PSD to frequency peaks at multiple frequencies, 
 with two of the most dominant peaks appearing at $f_m$ and the multiple of $f_m$ closest to $f_b$.
 In the snapshot images shown in subsequent sections, this collection of frequency peaks will be shown to coincide with a time-varying phase relationship between morphing and the underlying shedding process. 
 The morphing and timescales do not synchronize, and at different times the morphing near $s_{max}$ will be shown to alternately augment or inhibit the formation of the leading-edge vortex, and thereby modify the subsequent interaction with the trailing-edge vortex. These cycle-to-cycle variations in the vortex formation and interaction processes will be shown to lead to multiple timescales in the lift curve. 
 
 In the second category of behaviors when $f_m\not\approx f_b$, the morphing timescales are sufficiently fast compared with the underlying (unactuated) flow dynamics that the bulk vortex shedding processes and lift dynamics are not significantly affected. We refer to this behavior as the superposition regime because morphing actuation manifests itself in the lift dynamics as small-amplitude, high frequency oscillations superposed onto the (essentially) baseline lift variation. Cases belonging to this superposition regime are illustrated in figure \ref{fig:AOA15Regimes_FFT_New}h--i. This inability of the flow to respond to actuation at significantly disparate timescales from intrinsic flow timescales is the same as what was seen for $\alpha = 5^{\circ}$. 
 
  Before describing the detailed dynamics of each regime, it is worth clarifying the distinctions between the interactive and superposition regimes in more detail. We show in figure \ref{fig:AOA15Liftbehaviour} time histories of the lift dynamics at various wavespeed-wavenumber combinations that fall within either the interactive (figures \ref{fig:AOA15Liftbehaviour}a--c) or superposition (figure \ref{fig:AOA15Liftbehaviour}d) regimes. 
Note that there is a larger impact on the lift dynamics compared with the baseline case for lower morphing frequencies, $f_m=wn$. This impact reduces as $f_m$ increases, until it is nearly negligible for $(w,n) = (1,6)$ (figure \ref{fig:AOA15Liftbehaviour}d). This progression demonstrates a smooth transition between the interactive and superposition regimes as $f_m$ increases, where the flow becomes successively less receptive to actuation at very disparate timescales, $f_m \gg f_b$, until the primary flow dynamics are essentially those of the baseline (unactuated) case. Of the four subfigures, only \ref{fig:AOA15Liftbehaviour}d qualifies as superposition since it is the only one in which the effect of morphing appears as small amplitude, high frequency variations superposed onto the baseline lift dynamics.

Using these regime definitions, we present in figure \ref{fig:AOA15PM_NewII} performance maps that show the percentage increase in mean $C_l$ and $C_d$ compared to the unactuated flow. The different marker shapes represent the different regimes. The system exhibits lock-on when the morphing frequency $f_m=wm$ is near the baseline (unactuated) frequency $f_b$ if the wavenumber is not too high ($n \leq 4$). The interactive regime is observed for morphing frequencies smaller than or slightly above the baseline frequency. For sufficiently high morphing frequencies, the superposition regime is observed, where actuation has little effect in modulating the unactuated flow dynamics. Note that the receptivity of the system to morphing via either the lock-on or interactive regimes is stronger for lower wavenumbers. For example, for $n\ge5$ no lock-on is observed, and the interactive regime is observed for the highest frequencies when $n=1$. These facts reflect a secondary effect in addition to that of the morphing frequency: the actuation length-scale. When the morphing length-scales are significantly smaller than those of the bulk vortex-shedding processes, the system becomes less receptive to actuation and a superposition regime driven largely by the unactuated flow dynamics is observed.  

The largest mean lift improvements occur within the lock-on regime, whereas significant lift detriments are observed for the interactive regime and negligible changes occur within the superposition regime. Within lock-on, the maximal mean lift improvements are for low values of wavenumber ($n\le3$) where morphing lengthscales are of the same order as the primary vortical structures formed at the leading and trailing edges. Moreover, the highest performing cases $(w,n)=(0.6,1),$ $(0.3,2)$ are at a fixed morphing frequency of $f_m=0.6$. Note that lock-on is not exclusively beneficial to lift; e.g., for $n=4$ there is a mean lift reduction. A similar trend is seen for mean drag as in the lower angle of attack, with drag detriments coinciding for parameters yielding lift benefits, but with the drag increases being significantly smaller than lift increases. We therefore continue to focus on the effect of morphing on lift at this higher angle of attack. In subsequent subsections, we investigate each regime in detail to explore why the interactive regime is detrimental to lift, why within the lock-on regime certain parameters are either advantageous or disadvantageous for lift, and why there is a minimal change to aerodynamic performance in the superposition regime.  

To frame these detailed investigations, we show in figure \ref{fig:LockOnAdvection} a $t$-$x$ diagram of the local pressure minimum near the airfoil. In the analogous figure at the lower angle of attack of $5^\circ$, the pressure minimum was induced exclusively by morphing and advected at a local flow speed that was agnostic to the morphing wavespeed. For this higher angle of attack, the local pressure minimum is a signature of the leading-edge vortex. Interestingly, for the morphing parameters that lead to lock-on (indicated by the collection of blue curves), the advection speed is again largely constant despite modest changes in wavespeed and morphing frequency. This near-constant advection speed suggests a similar process as for the lower angle of attack: morphing modulates the vortical structures that form due to the unsteady shear layer (c.f., the different frequency peaks for the lift dynamics within the lock-on regime in figure \ref{fig:AOA15Regimes_FFT_New}d--g), but the subsequent advection of those structures after they are formed is unaffected by the morphing parameters. The near-constant advection speed is approximately 0.41. This result may appear counter-intuitive because of its difference from the 0.6 value one might expect as the optimal morphing frequency $f_m=0.6$ from the performance map in figure \ref{fig:AOA15PM_NewII}. We will clarify the reason for this advection speed using detailed snapshot information in subsequent subsections. 

In the interactive regime, there is no longer a uniform timescale that governs the advection of the flow structures induced by morphing. For example, when $(w,n) = (0.1, 2)$ there is no clear advection velocity and for $(w,n) \equiv (1.2, 1)$ (not shown in the performance map), the advection velocity is higher than the other cases in the advection velocity plot.
This variation suggests that, unlike at the lower angle of attack, the flow structures induced by morphing can be qualitatively distinct, yielding different local advection processes.
This result is intuitive because at this higher angle of attack of $15^{\circ}$, the pressure minimum is associated with the LEV. Thus, for cases where the morphing parameters are not synchronized with vortex shedding, changes in the LEV attributes could arise from complex phenomena including modification of the LEV formation process, interactions with a previous/subsequent LEV and/or the TEV, distance of the LEV from the airfoil surface and the effect of the velocity boundary condition due to morphing downstream of $s_{max}$, etc. These processes produce distinct and non-uniform advection speeds for the local pressure minimum associated with the LEV in figure \ref{fig:LockOnAdvection}.

In the superposition regime, the local pressure minimum advects in essentially 
the same manner as for the baseline case, consistent with observations about this regime made above. 

To better address the questions posed above regarding mean lift and the advection of the local pressure minimum associated with the leading-edge vortex, we utilize detailed snapshot information of the interactions between morphing and vortex shedding for each of the regimes separately below.

\subsection{Interactive Regime}
\label{sec:Interactive}
\begin{figure}
        \centering
        \includegraphics[width =0.8\textwidth]{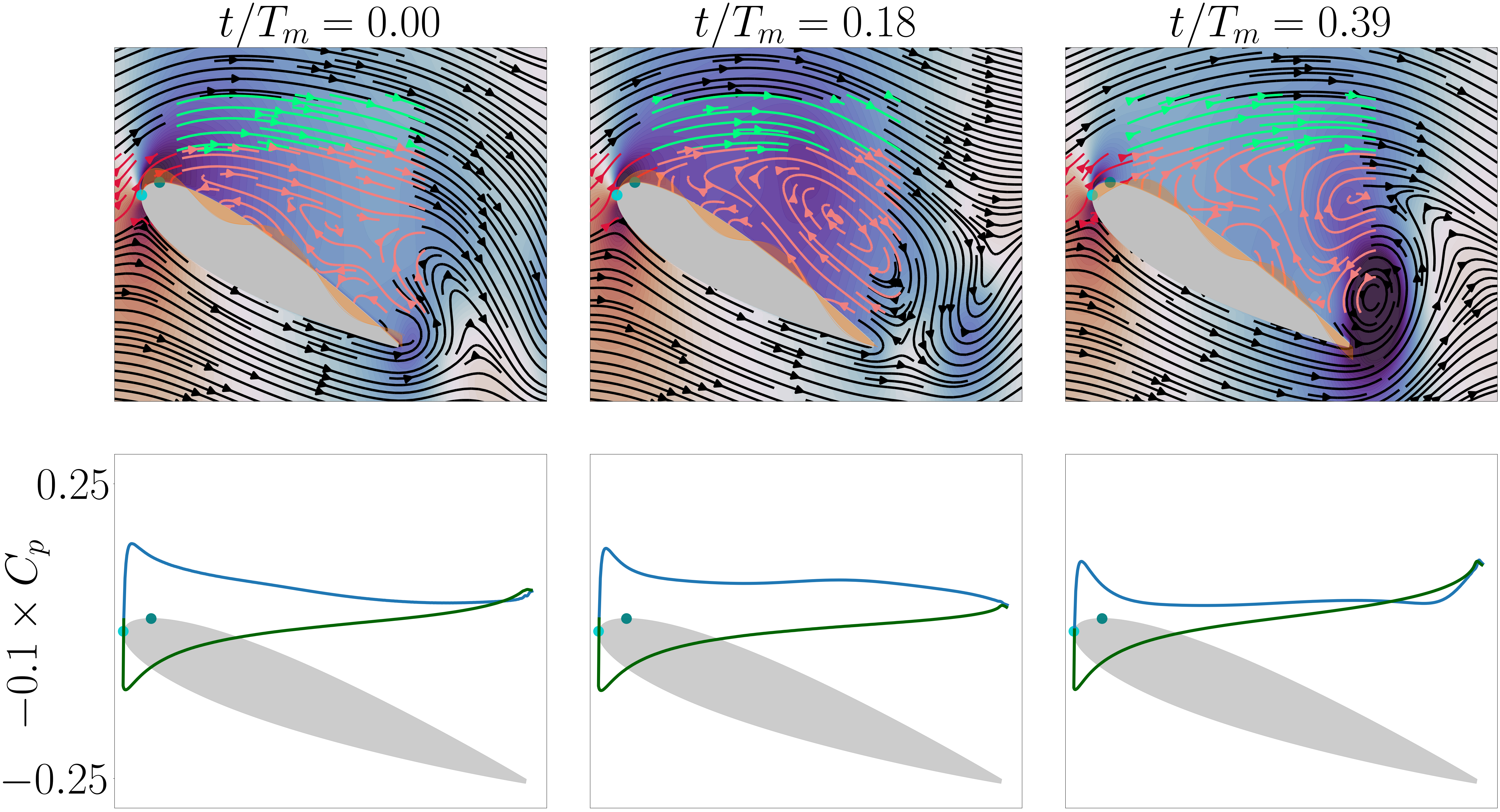}
    \vspace{0.1in}
        \includegraphics[width =0.8\textwidth]{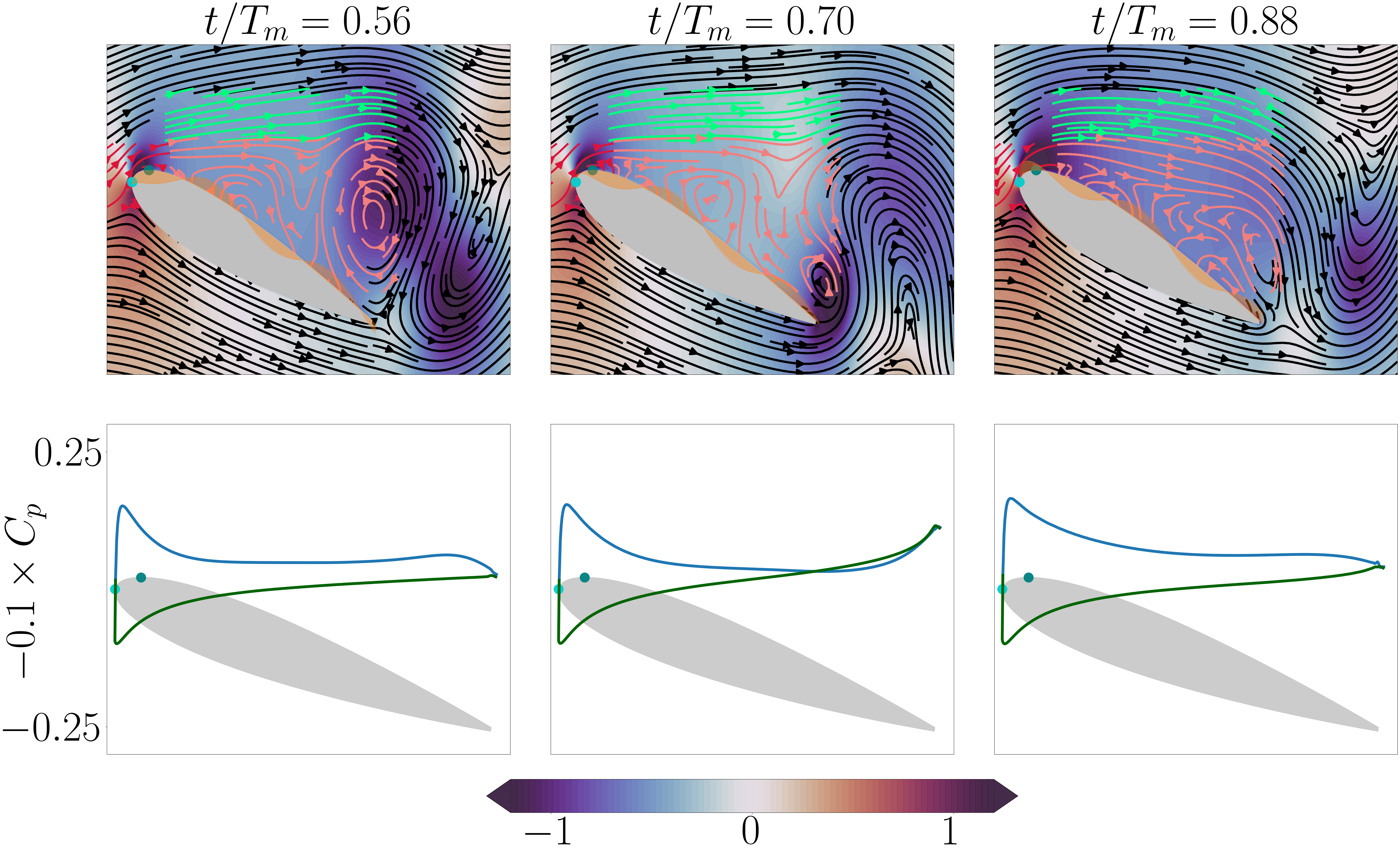}
    \caption{Analog of figure \ref{fig:AOA15BL}, $(w,n) = (0.1,2)$. Snapshots are at the instances indicated by the markers in \ref{fig:AOA15Regimes_FFT_New} (b). Note the additional snapshots
    utilized here to show how the vortex shedding cycle changes over a single morphing period.}
    \label{fig:AOA15n2c0.1}
\end{figure}

We now consider the interplay between surface morphing and the underlying vortex-shedding dynamics in the interactive regime. These interactions are most pronounced for small morphing frequencies $f_m=wn$ relative to the baseline (unactuated) frequency $f_b$, so we consider one of these cases with $(w,n)=(0.1, 2)$ for clarity of presentation. Extensions to other parameters are drawn at the end of this subsection.

Figure \ref{fig:AOA15n2c0.1} shows snapshots of the coefficient of pressure, $C_p$, in the flow-field and on the airfoil surface over one morphing cycle.
Note that since $f_m<f_b/3$, a single morphing period contains just over three vortex-shedding cycles in the baseline case. 
The description below also leverages information from the lift dynamics; c.f., figure \ref{fig:AOA15Regimes_FFT_New}b. At $t/T_m\approx 0$, the morphing velocity near $s_{max}$ is positive which increases the flow acceleration and nearby streamline convexity, and creates a more significant low-pressure zone than in the baseline case. This low-pressure region advects downstream to $t/T_m=0.18$ and a leading-edge vortex forms as indicated by the salmon streamlines. This LEV formation leads to an increase in lift as in the baseline case. As this LEV continues to form and advects near the trailing edge, a separate TEV forms and reaches near-maximal strength by $t/T_m=0.39$, leading to a local lift minimum. This set of processes---formation and advection of a LEV and its interaction with a subsequent TEV mark an analogous process to that seen in the unactuated case. At the same time, the presence of morphing over this shedding cycle introduces key (and lift-detrimental) differences. Most notably, at $t/T_m\approx0.39$ when the TEV is at its near maximum, the LEV has advected downstream of the airfoil and the negative morphing velocity near $s_{max}$ mitigates the low-pressure zone that would otherwise have formed in the baseline case to offset the lift minimum. (The detrimental effect of negative morphing near $s_{max}$ can also be seen by the low curvature of the green streamlines). As such, the minimal lift value is significantly lower for this set of morphing parameters than in the unactuated case.

Because of the lack of synchronization between morphing and the underlying vortex-shedding processes, the phase relationship between morphing and vortex shedding changes from one shedding cycle to the next. For example, the negative morphing velocity near $s_{max}$ at $t/T_m,\approx0.39$ mitigates the formation of the subsequent LEV compared with either $t/T_m\approx 0$ or the analogous part of the vortex-shedding cycle in the baseline case. As such, the smaller LEV appears further downstream than usual by $t/T_m\approx0.56$, and moreover the negative morphing velocity still present across $s_{max}$ continues to mitigate the low-pressure zone that forms near the leading edge. Thus, while the presence of the LEV leads to a lift improvement at that time instance, the benefit is significantly lessened from the prior vortex shedding cycle or from the baseline case. As the LEV continues to advect downstream, a TEV forms. Again, the negative morphing velocity near $s_{max}$ at this instance means that the low-pressure zone near the leading edge is less effective at mitigating the local lift minimum, and the minimal lift is significantly lower than in the baseline case.

Finally, in the third vortex-shedding cycle, the morphing velocity becomes positive near $s_{max}$, and the associated low-pressure zone and downstream advection of the lift-detrimental TEV at $t/T_m\approx0.88$ result in a local lift maximum. These three vortex-shedding cycles demonstrate that in the interactive regime, the same qualitative vortex-shedding process---formation of an LEV and an associated lift maximum and advection of that LEV/formation of a TEV and an associated lift minimum---persist. At the same time, the asynchrony between morphing and the underlying flow processes means that the specific strength, roll-up location, and advection speed of the LEV as well as the formation and subsequent interaction of the TEV vary across each shedding cycle. (In fact, note that the dynamics are not periodic over a morphing period: the lift dynamics from $t/T_m\in[0,1]$ are noticeably distinct from those over $t/T_m\in[1,2]$ in figure \ref{fig:AOA15Regimes_FFT_New}b). Altogether, these variations create broadband frequency signatures in the lift dynamics that are ultimately detrimental to lift (c.f., the performance map in figure \ref{fig:AOA15PM_NewII}). 

The modulating effect that morphing has on shedding within this interactive regime affects the advection of the local pressure minimum near the airfoil; c.f., figure \ref{fig:LockOnAdvection}. For example, for $(w,n) = (0.1, 2)$, the advection speed varies as the pressure minimum associated with the LEV traverses downstream. Moreover, the varying slope of the line towards the trailing edge coincides with TEV roll-up near $t/T_m\approx 0.39$. 

Note that while these observations were for the case in which there were several shedding cycles per morphing period, the underlying features hold for the other cases studied within this regime: there is the same qualitative shedding dynamics with cycle-to-cycle variations due to the de-synchronized morphing and shedding interplay. De-synchronization between morphing and shedding 
implies that over certain morphing cycles, morphing would impede formation of the LEV. This weaker LEV would less effectively mitigate the deleterious effects that the TEV has on lift, which suggests a mechanism through which the interactive regime is detrimental to mean lift.

Commensurate with this change in shedding behavior is a change in the advection properties of the LEV and associated local pressure minimum near the airfoil. For example, for $(w,n) = (1.2, 1)$ (in the interactive regime, not shown in performance map), the advection velocity is higher than for the other cases from the figure. For this wavespeed-wavenumber combination, the morphing frequency is faster than the baseline shedding frequency, resulting in two LEVs shed per cycle rather than one. These smaller LEVs form more quickly and advect faster downstream. The faster advection could be influenced by the interaction of these closer LEVs and their distance from the airfoil, among other factors. 

\subsection{Lock-on Regime}
\label{sec:Lockon}
\begin{figure}
    \centering
        \includegraphics[width =1\textwidth]{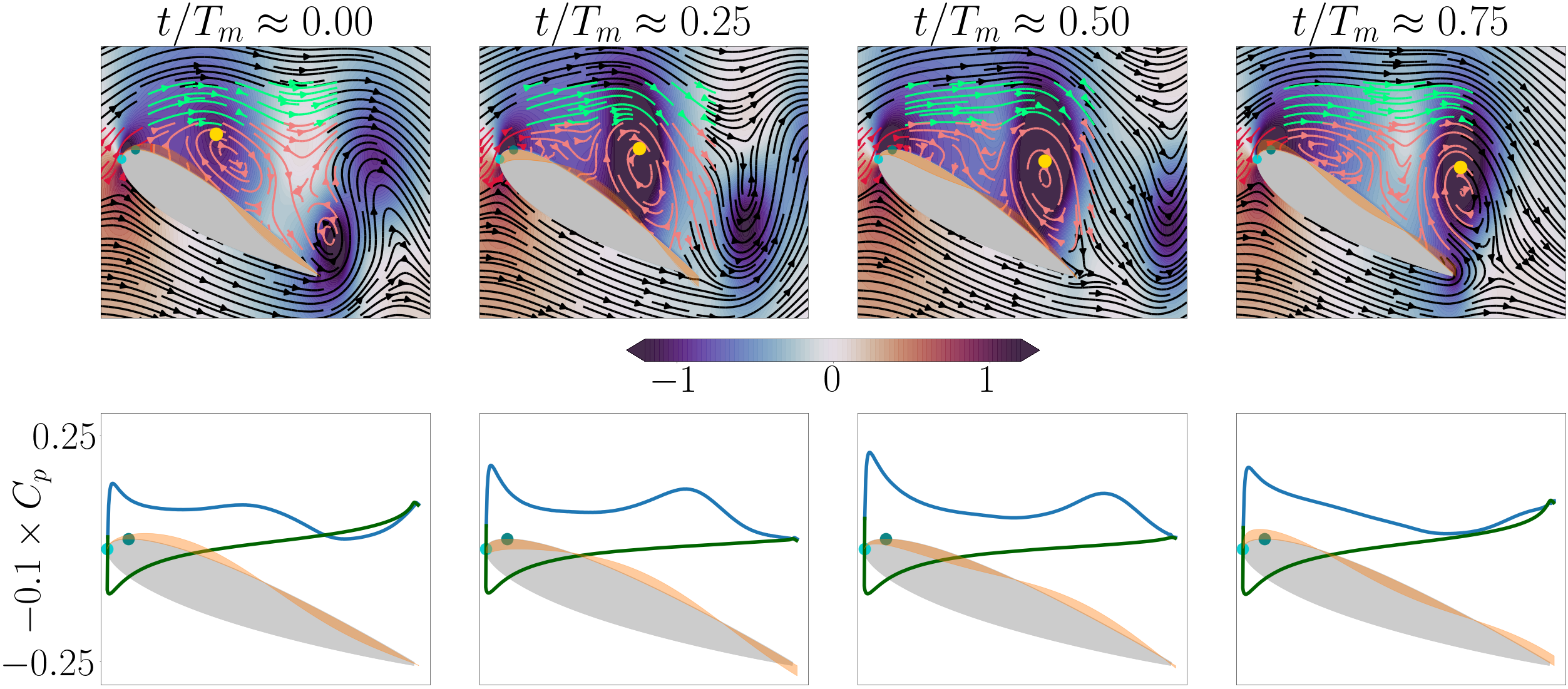}
    \caption{Analog of figure \ref{fig:AOA15BL}, $(w,n) = (0.6,1)$.}
    \label{fig:AOA15n1c0.6}
\end{figure}
\begin{figure}
    \centering
        \includegraphics[width =1\textwidth]{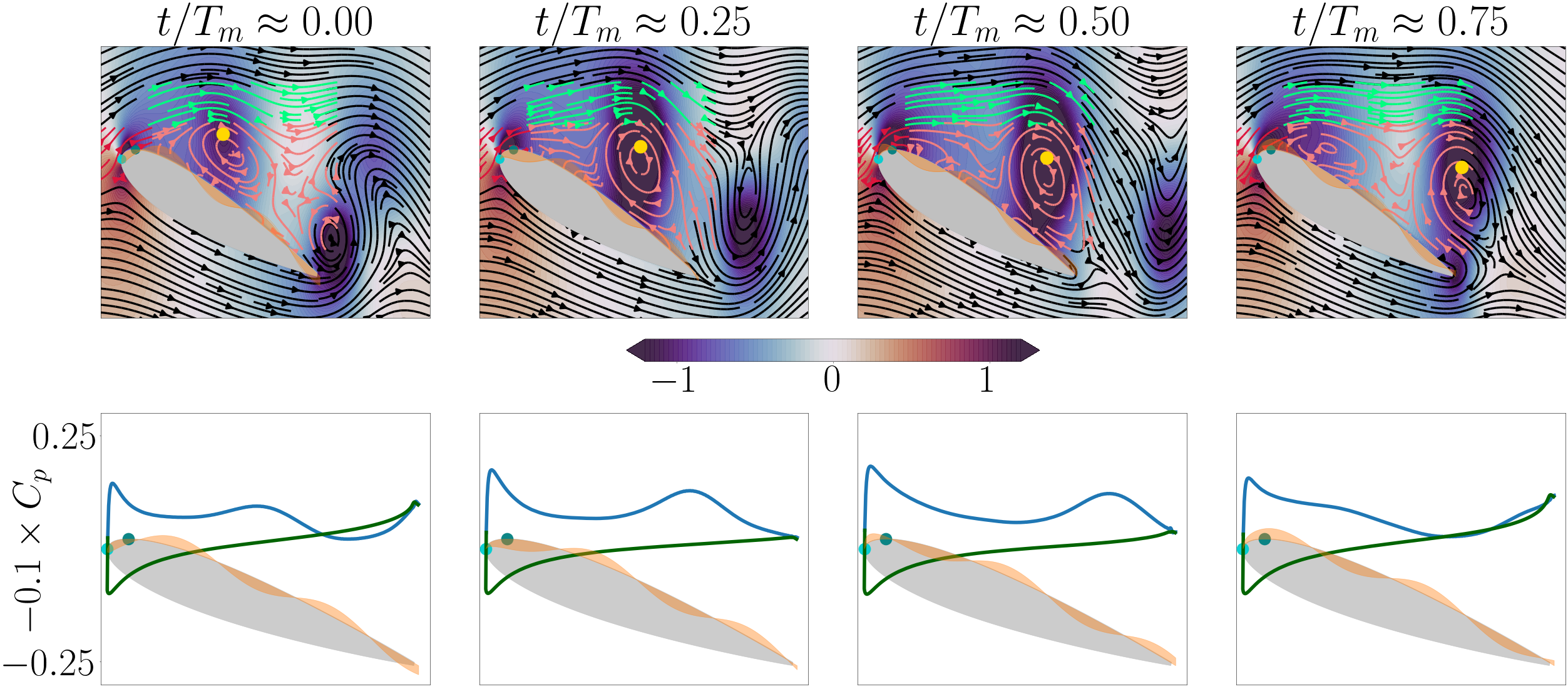}
    \caption{Analog of figure \ref{fig:AOA15BL}, $(w,n) = (0.3,2)$}
    \label{fig:AOA15n2c0.3}
\end{figure}

\begin{figure}
  \centering
   \includegraphics[width=0.95\textwidth]{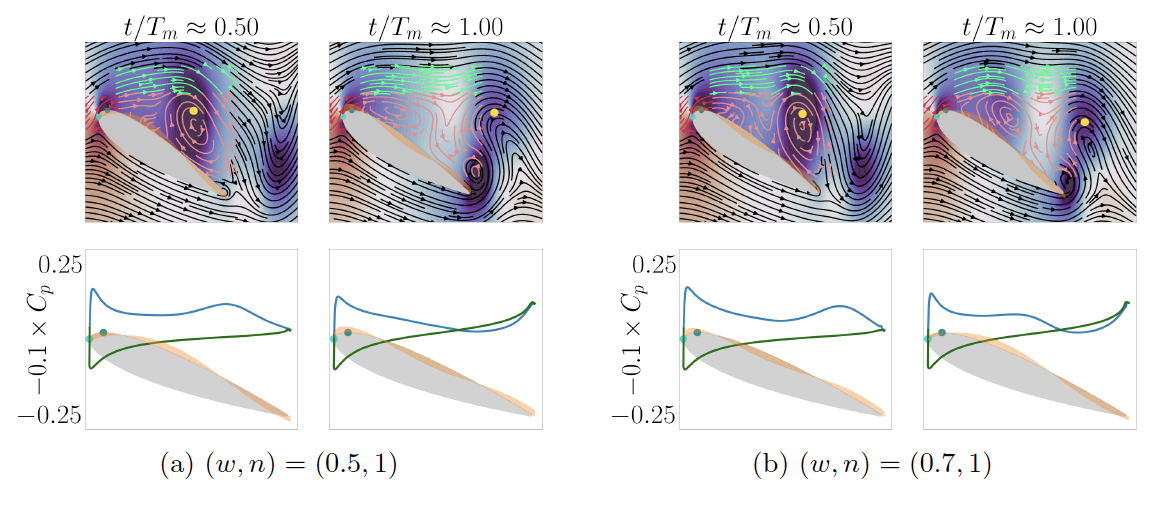}
     \caption{Analog of figure \ref{fig:AOA15BL}, $n=1$, and varied wavespeeds, $w$.}
     \label{fig:LockOnComparisonII}
\end{figure}

When $f_m \approx f_b$ (provided the wavenumber $n$ is not too large), the lock-on regime occurs where the lift dynamics synchronize with the morphing frequency and there is no remaining signature of the dynamics at the baseline (unactuated) frequency of $f_b$; c.f., figure \ref{fig:AOA15Regimes_FFT_New}. We show in this subsection that these lift dynamics reflect sychronization of morphing and vortex shedding. To clarify how morphing and vortex shedding interact in this synchronized state, and when this synchrony is beneficial for mean lift, we consider snapshots for various morphing parameters within the lock-on regime.

At $\alpha=15^{\circ}$, maximum mean lift is achieved at $(w,n) \equiv (0.6,1)$. Figure \ref{fig:AOA15n1c0.6} shows the vortex-shedding process over one morphing period for this $(w,n)$ combination. The temporal variation over one morphing cycle is shown in figure \ref{fig:AOA15Regimes_FFT_New}d, where the markers correspond to the snapshot instances in figure \ref{fig:AOA15n1c0.6}. Qualitatively, the snapshots show the same LEV-TEV shedding cycle as in the baseline (unactuated) case. However, there are key differences in the presence of morphing that are beneficial to lift.

At $t/T_m \approx 0$, low (lift-producing) pressure near the leading edge (and associated curvature of the green streamlines near $s_{max}$) is aided by the positive morphing velocity near $s_{max}$ from prior time instances. This process marks the early formation stages of the lift-beneficial LEV, and is in contrast to the baseline case, where these lift-producing effects are only  due to flow acceleration around the rounded leading edge. In addition, as the LEV advects and grows over $t/T_m=0.25-0.5$, the positive morphing velocity at $s_{max}$ and the favorable boundary conditions in the vicinity of the upward and downward pointing streamlines contribute to this growth, and the LEV ultimately has a significantly lower (lift-producing) associated pressure than in the unactuated case. When the LEV advects close to the trailing edge at $t/T_m \approx 0.75$, TEV formation starts to occur. However, the lift-reducing effects of the TEV are mitigated by partial LEV formation near $s_{max}$. Between $t/T_m \approx 0.75$ and $t/T_m \approx 0$ (i.e., $t/T_m \approx 1$), not only the TEV but also the LEV grows in strength and the drop in lift is less than that of the baseline flow, where only partial formation of the LEV would have occurred. Also, the morphing velocity near $s_{max}$ at $t/T_m \approx 0.75$ is near maximum which means that maximum morphing velocity would occur around the instance of when the LEV has maximum strength. Altogether, by aiding the formation of the LEV, morphing at this $(w,n)$ combination aids lift both because of the associated drop in pressure and the mitigation of lift reduction associated with the TEV.

To better understand the factors influencing variation in mean lift within the lock-on regime, we compare other $(w,n)$ combinations. We first investigate the role of wavenumber and consider $(w,n) = (0.3,2)$, which has the same $f_m$ as $(w,n) = (0.6,1)$. The $C_l$ variation with markers corresponding to the snapshots in figure \ref{fig:AOA15n2c0.3} is shown in figure \ref{fig:AOA15Regimes_FFT_New}e.
The $C_l$ variation as well as the snapshots look qualitatively similar to those at $(w,n) = (0.6,1)$. The differences are in the boundary conditions on the airfoil surface relative to the LEV. Comparing the snapshots at $t/T_m \approx 0$  for the two cases, it can be seen that the upward and downward pointing streamlines of the LEV are not as comprehensively aided by the morphing velocity profile when $(w,n) = (0.3,2)$ (for example, at $t/T_m =0.25, 0.5$ and $0.75$  the upward pointing streamlines of the LEV interface with negative morphing velocity). 
As a result, the LEV is narrower, weaker and some distance away from the airfoil. This makes the lift improvement lower than that of $n=1$.

We now consider the effect of wavespeed and investigate $w=0.5$ and $0.7$ at a fixed wavenumber of $n=1$. The corresponding $C_l$ variations are shown in figure \ref{fig:AOA15Regimes_FFT_New}f and g. Note from these lift curves that for $w=0.5$ the maximum (minimum) lift 
value is significantly higher (lower) than in the unactuated case. By contrast, for $w=0.7$ the lift maximum is only slightly larger than in the unactuated case, but the resulting mean lift is actually higher than for $w=0.5$ (c.f., figure \ref{fig:AOA15PM_NewII}) because the lift minimum is comparable to that of the unactuated case. This outcome suggests that the synchrony between morphing and vortex shedding observed in the lock-on regime results in two competing effects: augmenting maximum lift via favorable formation of the LEV and mitigating minimum lift that occurs through formation of the TEV. 

To clarify the manifestation of these effects within lock-on, we show in figure \ref{fig:LockOnComparisonII} snapshots at the instances of maximum and minimum lift for the two wavespeeds. For both cases at the instance of highest lift (near $t/T_m\approx 0.5$), the morphing velocity is not entirely advantageous to LEV formation 
(e.g., for $(w,n)=(0.5,1)$ the morphing velocity peak lags behind the upward swirling salmon streamlines associated with the LEV and for $(w,n)=(0.7,1)$, it leads). 
In analogy with $(w,n)=(0.3,2)$, this offset leads to a weaker LEV than for the highest performing case of $(w,n)=(0.6,1)$.  
Another effect of varying $f_m$ is the duration over which morphing velocity at $s_{max}$ is positive, and the impact this duration has on streamline curvature. Since the duration is longer for $w=0.5$, the convex (lift-beneficial) and concave (lift-detrimental) portions of the streamlines have larger spatial wavelengths. The larger convex portion of the streamlines is evident from the extent of the suction peak in the airfoil $C_p$ distribution at $t/T_m \approx 0.5$, and the larger concave portion is visible from the separation between consecutive LEVs at $t/T_m \approx 1$ (for $w=0.5$ the new LEV is yet to be formed).
The extended streamline convexity is associated with increased lift (evident in the higher magnitude and duration of lift increase above the baseline case for $w=0.5$, c.f. figure \ref{fig:AOA15Liftbehaviour}). However, for these morphing parameters the benefits are outweighed by the extended streamline concavity that is compounded by the coincident occurrence with the TEV formation process. The coincidence of the concave streamlines and the TEV formation are associated with a delayed LEV formation. This delay can be seen by comparing the snapshots at $t/T_m \approx 1$ for the two wavespeeds. At $w=0.7$, the new LEV roll-up is advanced and thus helps mitigate the lift detriments resulting from the TEV. On the contrary, the new LEV is hardly formed at this instance for $w=0.5$. The delayed LEV formation explains the drop in lift below the baseline value for $w=0.5$ and improvement in minimum lift for $w=0.7$.

% \begin{figure}
%     \centering
%         \includegraphics[width =1\textwidth]{AOA15_PGrid1_0.6.png}
%     \caption{Analog of figure \ref{fig:AOA15BL}, $(w,n) = (0.6,1)$.}
%     \label{fig:AOA15n1c0.6}
% \end{figure}
% \begin{figure}
%     \centering
%         \includegraphics[width =1\textwidth]{AOA15_PGrid2_0.3.png}
%     \caption{Analog of figure \ref{fig:AOA15BL}, $(w,n) = (0.3,2)$}
%     \label{fig:AOA15n2c0.3}
% \end{figure}

% \begin{figure}
%   \centering
%   \includegraphics[width=0.95\textwidth]{AOA15_LockOnComparison.png}
%      \caption{Analog of figure \ref{fig:AOA15BL}, $n=1$, and varied wavespeeds, $w$.}
%      \label{fig:LockOnComparisonII}
% \end{figure}

\begin{figure}
    \centering
        \includegraphics[width =1\textwidth]{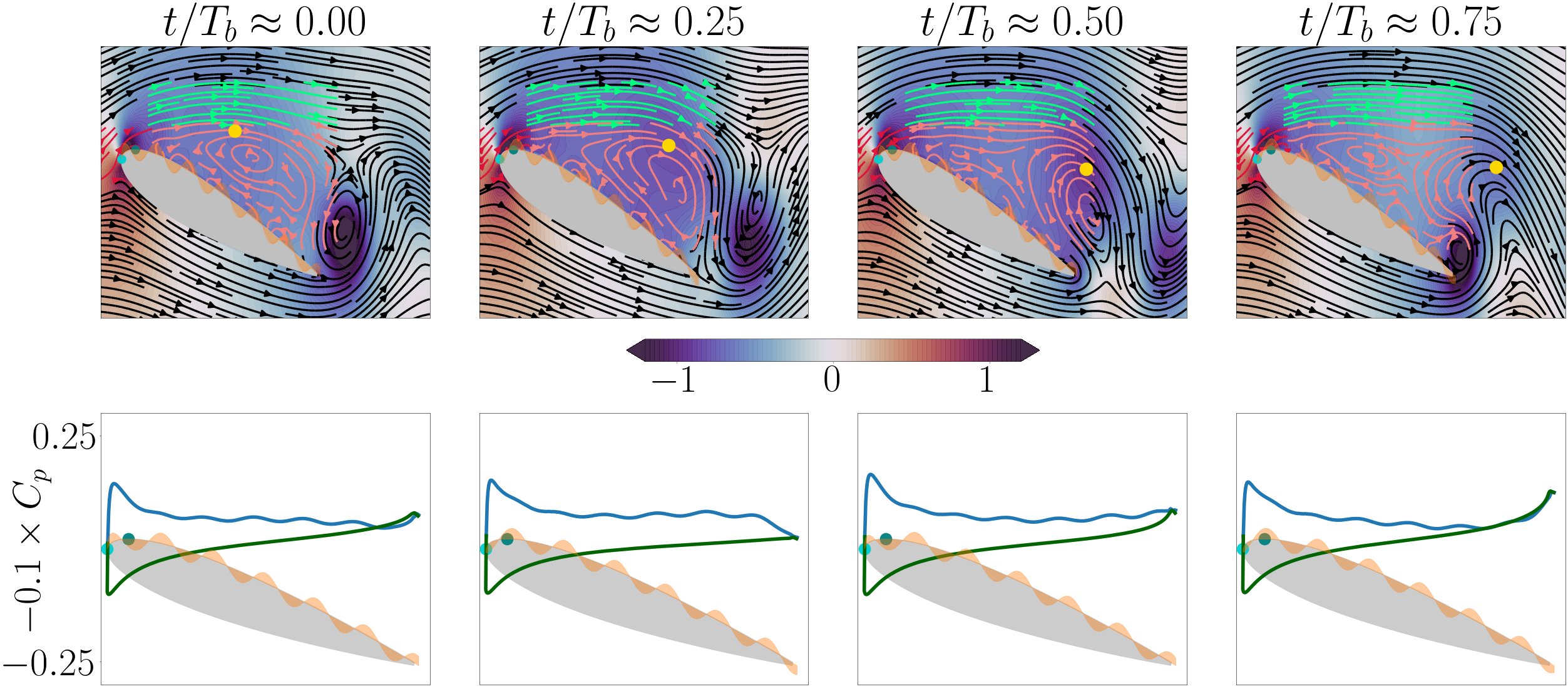}
    \caption{Analog of figure \ref{fig:AOA15BL}, $(w,n) = (0.7,6)$.} 
    \label{fig:AOA15n6c0.7}
\end{figure}

\subsection{Superposition Regime}
\label{sec:Superposition}

When the morphing frequency $f_m$ is too large to affect the intrinsic vortex shedding process, the lift dynamics are largely unchanged. Figure \ref{fig:AOA15n6c0.7} shows snapshot images for an example %where 
of a high wavenumber $n$ resulting in high $f_m$. The morphing velocity at $s_{max}$ during LEV formation results in slight perturbations (visible as spatial fluctuations on the $C_p$ airfoil distribution images) to what is otherwise essentially the shedding process from the unactuated case; c.f., figure \ref{fig:AOA15BL}. The oscillations about the lift variation of the unactuated flow are due to the integrated effect of the pressure oscillations on the airfoil surface. As was seen for $\alpha = 5^{\circ}$, the pressure peaks coincide with clockwise streamlines in regions of attached flow and anti-clockwise streamlines in regions of reversed flow. The flow reversal here occurs in tandem with the upstream pointing streamlines of the LEV. Because the underlying (unactuated) shedding process is largely unaltered, the advection properties of the pressure minimum associated with the LEV are essentially unaltered from the unactuated case (c.f., figure \ref{fig:LockOnAdvection}).

\section{Conclusions}
In this article, we studied the effects of surface morphing at $Re = 1{,}000$ on an airfoil at angles of attack of $5^{\circ}$ and $15^{\circ}$. 
In both cases, it was shown that morphing could be utilized to change streamline curvature and hence the pressure distribution on the suction surface of the airfoil in a manner beneficial to mean lift. 
The pressure signature from the leading edge to the point of largest $y$-coordinate value, 
$s_{max}$, was shown 
to be driven by the morphing kinematics over that region. 
Morphing velocity in this region also led to clockwise streamlines which affected the curvature of the streamlines near $s_{max}$. The streamlines were shown to have maximum convexity when morphing velocity near $s_{max}$ was maximum. Streamline convexity was linked to reduced pressure and lift benefits. On the contrary, lack of streamline convexity which resulted from negative morphing velocity around $s_{max}$ was linked to increased pressure and lift detriments.
At the angle of attack of $5^{\circ}$, this phenomenon was shown to be the reason for temporal variation of lift while at $15^{\circ}$, the lift dynamics were a more complex interplay between the periodic morphing signature and the underlying vortex-shedding dynamics.

At $\alpha = 5^{\circ}$, the advection speed of the (lift-producing) suction peak created at $s_{max}$ was shown to be largely unaffected by morphing kinematics. First, a study at a fixed wavenumber $n=1$ was performed, and the maximum lift improvement was shown to occur when the morphing wavespeed matched this universal advection speed. This phenomenon was explained in terms of the morphing frequency encoded in terms of the wavespeed $w$ and wavenumber $n$ as $f_m=wn$: when the morphing period matched the duration for the lift-beneficial suction peak to travel from $s_{max}$ to the trailing edge, then this meant that a new lift producing suction peak could appear at $s_{max}$ just as the older one was advecting beyond the airfoil.  By contrast, for longer morphing periods the suction peak  advected downstream of the airfoil before the suction peak re-appeared at $s_{max}$, and for faster periods the magnitude of the beneficial pressure suction peak was reduced. Mean lift was maximized when these competing effects were balanced, when the morphing wavespeed matched the local advection speed of the pressure signature.

At this lower angle of attack of $\alpha=5^\circ$, maximum mean lift was found at wavespeed $w=0.5$ and wavenumber $n=2$. To probe the benefits of $n=2$ over $n=1$ and to study the effect of wavenumber at fixed wavespeed, comparisons were made between $n=1, 2$ and $3$ with the wavespeed fixed at $0.5$. With increasing wavenumber, the increasing morphing frequency led to more frequent formation of suction peaks at $s_{max}$, albeit of lower magnitude. The effect of multiple peaks was a reduction in maximum lift (each peak was weaker) and an increase in minimum lift (the persistence of a suction peak over all times mitigated the lowest part of the lift cycle). The ideal balance of these effects was found for $n=2$. Finally, for high values of wavenumber, the short wavelength and high frequency limited the extent to which morphing impacted the global flow-field and thereby the pressure distribution. The effect of morphing was limited to small wavelength, high frequency pressure peaks of negligible magnitude close to the airfoil surface.

The observations made at $\alpha= 5^\circ$ were used to understand the higher angle of attack setting of $\alpha=15^\circ$, where the unactuated flow exhibits limit-cycle dynamics involving alternate shedding between a leading- and trailing-edge vortex. The flow dynamics in the absence of morphing showed attributes similar to actuated flow at $\alpha = 5^{\circ}$, with the additional phenomenon of trailing edge vortex, TEV, formation. The advection of the suction peak associated with the leading edge vortex, LEV, close to the trailing edge was associated with formation of the TEV which reduced lift considerably. 
In the unactuated flow, the TEV formation coincided with partial formation of the LEV near $s_{max}$ which to an extent mitigated the lift detriments of the TEV owing to the suction peak associated with the LEV.
The influence of morphing was shown to be driven by the relationship between the morphing frequency, $f_m$, and frequency of the baseline (unactuated) shedding dynamics, $f_b$. When $f_m \approx f_b$ (for sufficiently low wavenumbers), lock-on occurred, where the shedding frequency shifted to morphing frequency. Similar to the low angle of attack setting, within this regime the advection speed of the suction peak (associated with the LEV) was found to be nearly identical for the different wavespeed-wavenumber combinations, with a value near 0.41. Maximum performance coincided with $f_m$ that led to the highest morphing velocity at $s_{max}$ during TEV formation, which was found to coincide with the greatest benefit to LEV formation. These morphing parameters were also shown to yield velocity boundary conditions that aided LEV formation throughout its advection from pinch-off to the trailing edge. As was found in the lower angle of attack setting, this maximally beneficial parameter set was associated with a morphing frequency that aligned with the advection speed of the pressure suction peak. 

By contrast, when $f_m$ and $f_b$ were distinct from one another or $n$ was sufficiently large, two categories of behavior were found to be possible. In the first classification were cases where $f_m>>f_b$, termed here the superposition regime. In this regime, the flow was largely unresponsive to the relatively fast morphing timescales, and the lift dynamics were largely unaffected, just as for $\alpha=5^\circ$. At this higher angle of attack, morphing appeared as a high frequency perturbation to the essentially unactuated lift dynamics in a manner that did not affect the mean. In the second classification were parameters for which lock on did not occur, but at the same time the morphing timescales were sufficiently close to the underlying shedding dynamics for the flow to respond. In this regime, termed here the interactive regime, there were broadband frequency dynamics that coincided with cycle-to-cycle variations in the interplay between morphing and the vortex-shedding dynamics. These variations were due to a phase difference between morphing and shedding: during some shedding cycles, the partial LEV formation was aided while for some others it was opposed. In the former cycles, there were lift benefits due to stronger LEVs that provided greater suction peaks and mitigated the deleterious effect of the TEV. By contrast, in the latter cycles, LEV formation was delayed and the suction peaks were smaller and the lift detriments of the TEV less effectively mitigated. These disadvantageous cycles led to an overall decrease in mean lift.

\begin{acknowledgments}
The authors acknowledge Mr. Kevin Triner for obtaining preliminary data and for help with early stages of the analysis process.
\end{acknowledgments}

%\nocite{*}

%\newpage

\appendix

\section{Grid convergence}
\label{appA}
\begin{figure}[ht]
    \centering
    \begin{subfigure}{.49\textwidth}
        \centering
        \includegraphics[width =1\linewidth]{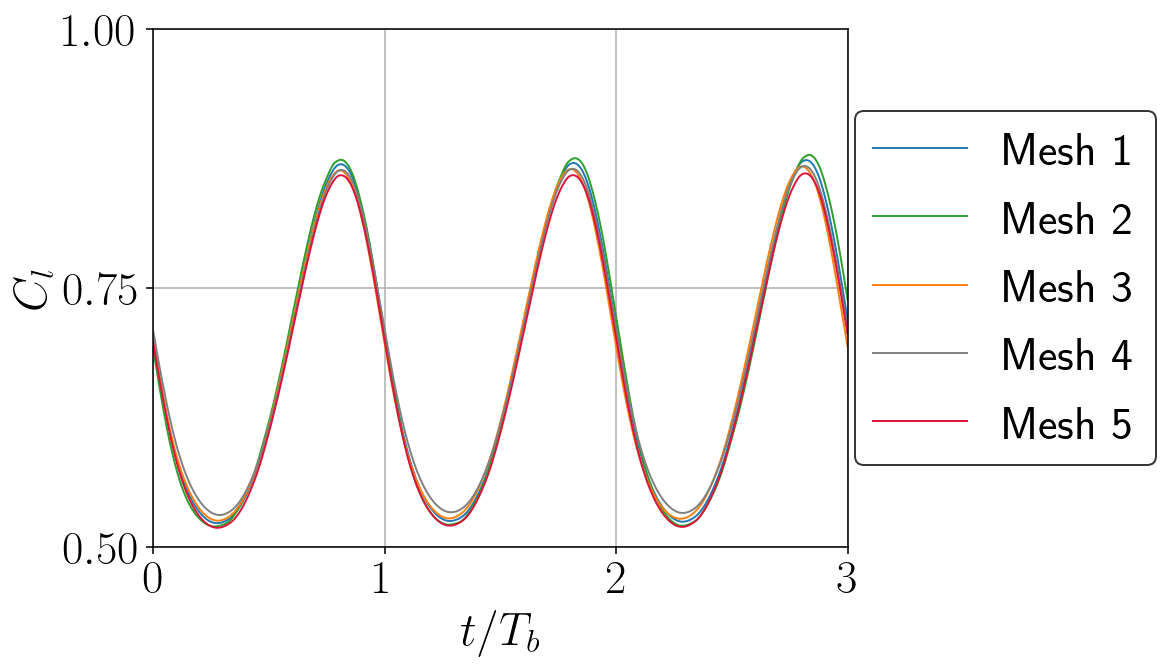}
        \label{fig:AOA15GridI}
    \end{subfigure}
    \begin{subfigure}{.49\textwidth}
        \centering
        \includegraphics[width =1\linewidth]{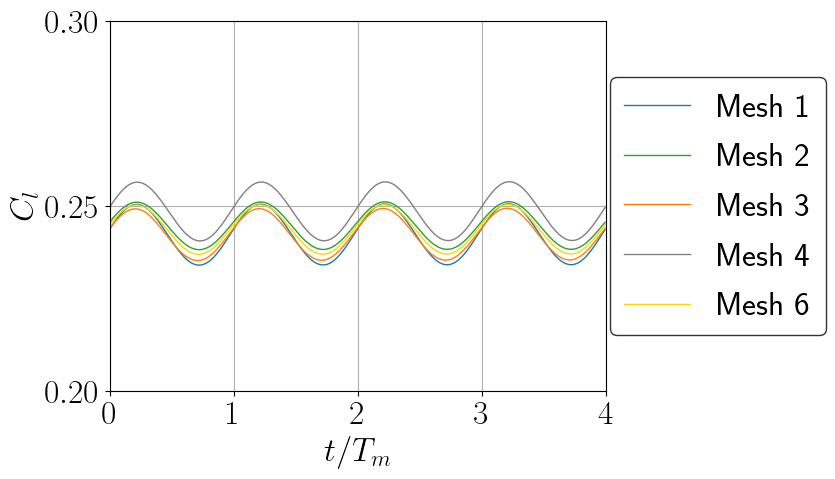}
        \label{fig:AOA15GridI}
    \end{subfigure}
    \caption{Grid convergence. Left: $\alpha = 15^{\circ}$, no morphing; Right: $\alpha = 5^{\circ}$, Morphing with highest $f_m$ considered in the article. }
    \label{fig:GridConv}
\end{figure}

The suitability of the simulation parameters used in this article  is justified through the comparisons given in tables \ref{table:AOA15Grid} and \ref{table:AOA5Grid} as well as the $C_l$ variations for angle of attack of $15^{\circ}$ without morphing and $5^{\circ}$ with morphing in figure \ref{fig:GridConv}. The grid convergence comparison includes the effect of grid spacing, total domain size (by increasing the grid levels of the multi-domain grid) and spacing of body points relative to grid spacing. The time step for each grid was chosen such that the Courant-Friedrich-Levy number is approximately $0.2$.
Grid $1$ is the one used for the simulations in this article. 
In each table, the last column is the percentage change in mean $C_l$ as compared to grid $1$. The quantity $\Delta s$ in the third to last column represents the spacing between the body points.
For $\alpha = 5^{\circ}$, domain independence was not tested since domain independence was established at $\alpha = 15^{\circ}$ where the vortical structures are stronger than at $5^{\circ}$.

\begin{table}[h]
\begin{center}
 \begin{tabular}{ |c|c|c|c|c|c|c|c|} 
 \hline
 Grid & $\Delta x$ & $\Delta t$ & Smallest sub-domain & Total domain &$\frac{\Delta s}{\Delta x}$& $C_l$  & $\% |\delta C_l|$
 \\ 
 \hline 
 1* & 0.00349 & 0.0004375 & [-0.5,2.5] $\times$ [-1.5,1.5]  & [-23, 25]  $\times$ [-24,24] &2 & 0.675 & N/A  \\ 

 2 & 0.005 & 0.0006 & [-0.5,2.5] $\times$ [-1.5,1.5] &  [-23, 25]  $\times$ [-24,24] &2  & 0.676 & 0.15  \\ 

3 & 0.0025 & 0.0003125 & [-0.5,2.5] $\times$ [-1.5,1.5] & [-23, 25]  $\times$ [-24,24] &2 & 0.674 & 0.15  \\ 

4 & 0.0025 & 0.0003125 & [-0.5,2.5] $\times$ [-1.5,1.5] & [-23, 25]  $\times$ [-24,24] & 1.5 & 0.678 & 0.44  \\ 

5 & 0.00349 & 0.0004375 & [-0.5,2.5] $\times$ [-1.5,1.5] & [-47, 49]  $\times$ [-48,48] & 2 & 0.668 & 1.04 \\

6 & 0.00349 & 0.0003 & [-0.5,2.5] $\times$ [-1.5,1.5] & [-23, 25]  $\times$ [-24,24] & 1.5 & N/A & N/A \\
\hline
\end{tabular}
\end{center}

 \caption{Grid convergence for $\alpha = 15^{\circ}$ without morphing}
\label{table:AOA15Grid}
\end{table}

\begin{table} [h]
\begin{center}
 \begin{tabular}{ |c|c|c|c|c|c|c|c|} 
 \hline
 Grid & $\Delta x$ & $\Delta t$ & Smallest sub-domain & Total domain &$\frac{\Delta s}{\Delta x}$& $C_l$  & $\% |\delta C_l|$
 \\ 
 \hline 
 1* & 0.00349 & 0.0004375 & [-0.5,2.5] $\times$ [-1.5,1.5]  & [-23, 25]  $\times$ [-24,24] &2 & 0.242 & N/A  \\ 

 2 & 0.005 & 0.0006 & [-0.5,2.5] $\times$ [-1.5,1.5] &  [-23, 25]  $\times$ [-24,24] &2  & 0.245 & 1.24  \\ 

3 & 0.0025 & 0.0003125 & [-0.5,2.5] $\times$ [-1.5,1.5] & [-23, 25]  $\times$ [-24,24] &2 & 0.242 & 0.0  \\ 

4 & 0.0025 & 0.0003125 & [-0.5,2.5] $\times$ [-1.5,1.5] & [-23, 25]  $\times$ [-24,24] & 1.5 & 0.249 & 2.9 \\  

5 & 0.00349 & 0.0004375 & [-0.5,2.5] $\times$ [-1.5,1.5] & [-47, 49]  $\times$ [-48,48] & 2 & N/A & N/A \\

6 & 0.00349 & 0.0003 & [-0.5,2.5] $\times$ [-1.5,1.5] & [-23, 25]  $\times$ [-24,24] & 1.5 & 0.244 & 0.8 \\
\hline
\end{tabular}
\end{center}
  \caption{Grid convergence for $\alpha = 5^{\circ}$ with $(w,n) \equiv (1, 6)$, the highest $f_m$ considered in this article.}
  \label{table:AOA5Grid}
\end{table}

%\bibliography{apssamp}% .
\newpage

\bibliography{SM_bib}% Produces the

\end{document}